\documentclass[fleqn,usenatbib]{mnras}

\usepackage{newtxtext,newtxmath}

\usepackage[T1]{fontenc}
\usepackage{ae,aecompl}
\usepackage{multirow}
\usepackage{units}

\usepackage{graphicx}	
\usepackage{listings}
\usepackage[separate-uncertainty=true,multi-part-units=single]{siunitx}
\usepackage{xcolor}
\usepackage{ulem}
\usepackage{afterpage}
\newcommand\T{\rule{0pt}{2.6ex}}		 
\newcommand\B{\rule[-1.2ex]{0pt}{0pt}} 
\newcommand{\hrefdoi}[2][ ]{\href{http://dx.doi.org/#2}{#1}}

\urldef{\footurlone}\url{ftp://gong2.nso.edu/TSERIES/v1f}
\urldef{\footurltwo}\url{ftp://ftp.seismo.nrcan.gc.ca/spaceweather/solar_flux/daily_flux_values/}
\urldef{\footurlthree}\url{https://gong2.nso.edu/fill.txt}
\urldef{\urlfour}\url{ftp://ftp.seismo.nrcan.gc.ca/spaceweather/solar_flux/daily_flux_values/}

\usepackage[hyphenbreaks]{breakurl}

\title[p-Mode Energy Supply Rates]{They Do Change After All: 25 Years of GONG Data Reveal Variation of p-Mode Energy Supply Rates}

\author[R. Kiefer \& A.-M. Broomhall]{
	Ren\'e Kiefer\thanks{E-mail: r.kiefer@warwick.ac.uk},
	Anne-Marie Broomhall,
	\\
	Centre for Fusion, Space, and Astrophysics, Department of Physics, University of Warwick, Coventry, CV4 7AL, United Kingdom
}

\date{Accepted XXX. Received YYY; in original form ZZZ}

\pubyear{2020}

\begin{document}
	\label{firstpage}
	\pagerange{\pageref{firstpage}--\pageref{lastpage}}
	\maketitle


\begin{abstract}
It has been shown over and over again that the parameters of solar p modes vary through the solar activity cycle: frequencies, amplitudes, lifetimes, energies. However, so far, the rates at which energy is supplied to the p modes have not been detected to be sensitive to the level of magnetic activity.
We set out to re-inspect their temporal behaviour over the course of the last two Schwabe cycles.
For this, we use Global Oscillation Network Group (GONG) p-mode parameter tables. We analyse the energy supply rates for modes of harmonic degrees $l=0\text{--}150$ and average over the azimuthal orders and, subsequently, over modes in different parameter ranges. This averaging greatly helps in reducing the noise in the data.
We find that energy supply rates are anti-correlated with the level of solar activity, for which we use the $F_{10.7}$ index as a proxy. Modes of different mode frequency and harmonic degrees show varying strengths of anti-correlation with the $F_{10.7}$ index, reaching as low as $r=-0.82$ for low frequency modes with $l=101\text{--}150$.
In this first dedicated study of solar p-mode energy supply rates in GONG data, we find that they do indeed vary through the solar cycle. Earlier investigations with data from other instruments were hindered by being limited to low harmonic degrees or by the data sets being too short. We provide tables of time-averaged energy supply rates for individual modes as well as for averages over disjunct frequency bins.
\end{abstract}


\section{Introduction}\label{sec:1}
Solar acoustic oscillations (p modes) are stochastically excited in the turbulent, outer convective layers of the Sun \citep[e.g.,][]{Balmforth1992, Rimmele1995, Houdek2015}. They are well described by a damped and stochastically forced harmonic oscillator in the time domain \cite[e.g.,][]{Anderson1990}. In turn, in the Fourier domain a single p-mode peak is well characterised by a Lorentzian profile with a frequency $\nu$, width $\Gamma$, and height $A$ (neglecting the observed mode asymmetry, e.g., \citealp{Nigam1998, Korzennik2017, Philidet2020}). Measurements of these mode parameters in the Fourier domain can be used to learn about the energetics of the oscillator in the time domain: the peak width $\Gamma$ is inversely proportional to the mode's lifetime and thus holds information about the damping. The product of peak height -- or amplitude -- $A$ and $\Gamma$, after proper scaling with the corresponding mode mass, encodes the energy $E$ that is stored in the respective p-mode oscillation \citep[e.g.,][]{Goldreich1994, Baudin2005}. Ultimately, the energy that is supplied to a mode per unit time $dE\slash dt$ is proportional to the product $\Gamma^2 \cdot A$ \citep[e.g.,][]{Goldreich1994}. This quantity is thus directly connected to the forcing of the mode \citep[e.g.,][]{Chaplin2003}. Possible temporal changes in the forcing function of the oscillator can thus be measured via time-resolved measurement of the quantity $\Gamma^2 \cdot A$.

Magnetic activity affects p modes in various ways: frequencies of modes below the acoustic cut-off \citep{Balmforth1990}, i.e., modes that are essentially trapped in the solar interior, are correlated with the level of magnetic activity: frequencies are highest during times of strong solar activity \citep[e.g.,][]{Woodard1985, Elsworth1990, Libbrecht1990, Jimenez-Reyes1998, Chaplin2001, Salabert2015a, Tripathy2015, Broomhall2017}. For oscillations above the acoustic cut-off -- these are waves that escape into the solar atmosphere \citep{Balmforth1990, Fossat1992, Vorontsov1998} -- frequencies turn anti-correlated with the activity cycle \citep[see, e.g., ][]{Woodard1991, Howe2008a, Rhodes2010}. The magnitude of activity-related shifts of mode frequencies depend on both mode frequency and harmonic degree $l$ (see \citealt{Basu2016} and references therein).

On the energetic side, p modes are increasingly damped with more activity on the surface, which leads to broader peak widths, as the two are inversely related to each other (e.g., \citealp{Jefferies1991, Komm2000a, Chaplin2000, Jimenez2002, JimenezReyes2003, JimenezReyes2004, Salabert2007, Burtseva2009, Broomhall2015, Kiefer2018a} and see \citealt{Broomhall2014} for a more extensive list of references on the matter). The observed changes along the solar cycle are measured to be between \unit[10--20]\si{\percent} depending on mode frequency and harmonic degree \citep[e.g.,][]{Kiefer2018a}. Mode amplitudes are anti-correlated with activity \citep[e.g.,][]{Palle1990, Elsworth1993, Chaplin2000, Komm2000a, Jimenez2002, JimenezReyes2003, JimenezReyes2004, Broomhall2015, Kiefer2018a}. Again, the magnitude of these changes is somewhat dependent on mode frequency and harmonic degree \citep[e.g.,][]{Kiefer2018a} but are typically between \unit[10--25]\si{\percent} over the solar cycle. As mode amplitudes can be converted into mode energies, these are also anti-correlated with the level of activity.

The energy supply rate (or excitation rate, forcing rate, energy dissipation rate through the oscillator) of the p modes has thus far been assumed to be constant: This has been explained on the simple reason that changes in damping are of the same magnitude as the accompanying changes in mode energies. Thus, changes in mode damping are sufficient to explain the entire observed phenomenology \citep{Chaplin2000}: modes get broader and their amplitudes get smaller. In turn, the increased damping explains the lower mode energies during phases of higher activity without having to propose a variation to the forcing function. For more details on how mode damping, excitation, and forcing are related under the assumption of a harmonic oscillator, we refer to \cite{Chaplin2000}, in particular their Section 2.

In addition, measurements of the energy supply rates -- via the compound quantity $\Gamma^2 \cdot A$ -- have repeatedly returned constancy through the solar activity cycle \citep{Chaplin2000, JimenezReyes2004, JimenezReyes2003, Salabert2007, Broomhall2015}. However, all of these studies investigated only low harmonic degrees $l\leq3$ from disk-integrated Sun-as-a-star data.

The energy supply rate of p modes is intimately connected with the spatial and temporal properties of convection \citep[e.g.,][]{Samadi2003a}. The temporal correlation of convective eddies is crucial to the theoretical understanding and modelling of mode energy supply rates \citep[e.g.,][]{Houdek2010}. Different analytical descriptions of the eddy-time-correlation can lead to vastly different values in the predicted energy supply rates \citep[e.g.,][]{Belkacem2010a}. Given the observed changes of near-surface convection properties through the solar cycle, e.g., a decrease in the size of granules \citep{Macris1984,Muller1988} and decreasing granular contrast with increasing level of magnetic activity \citep{Muller2007}, it can be expected that the energy supply rates are in fact not entirely constant through the solar activity cycle.

Short term departures from constancy have already been detected for disk-integrated helioseismic observations before: \cite{Chaplin2003} report on a roughly \SI{100}{\day} long augmentation of the mode forcing during 1998 in Birmingham Solar Oscillation Network (BiSON) data. 

A hint in GONG data towards activity-related changes of p-mode energy supply rates was found by \cite{Komm2000b} but has never been substantiated. Also, \cite{Komm2000} noted that energy supply rates decrease with increasing activity, but the detected change amounted to a mere \SI{-4.4(43)}{\percent} for the average over modes with $15\leq l\leq 95$ and $\SI{1.6}{\milli\hertz}\leq\nu\leq\SI{3.1}{\milli\hertz}$. Thus, no long-term variation of the mode energy supply rates that is {\mbox{(anti-)correlated}} with the level of magnetic activity has been conclusively confirmed as yet.

In the following we describe the data we used in this study (Section~\ref{sec:2}) and how we prepared and corrected these data (Section~\ref{sec:3}). The results from our analyses are presented in Section~\ref{sec:4}. A detailed discussion follows in Section~\ref{sec:6}, closing with the conclusions we draw from this study and its findings in Section~\ref{sec:7}.

\begin{figure}
\centering	\includegraphics[width=\linewidth]{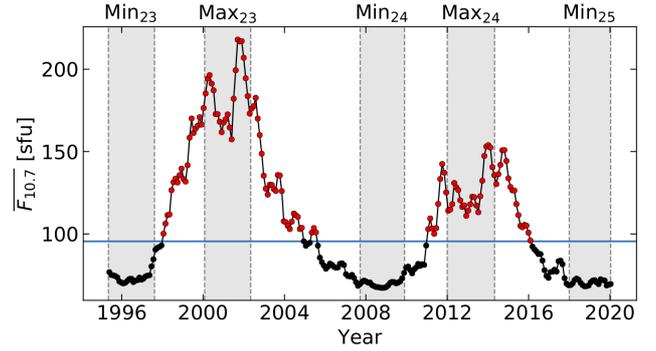}\caption{$\overline{F_{10.7}}$ solar radio flux during the GONG operation. The blue horizontal line indicates the median value. Times of activity maxima and minima are highlighted by the grey areas. Individual data points are averages of daily $F_{10.7}$ values over the 108 days of the GONG datasets.}\label{fig:f107}
\end{figure}

\begin{table}
	\centering
	\begin{center}	\caption{Start and end dates of solar activity extrema, mean $\overline{F_{10.7}}$ index during, and number of GONG datasets during these times.}\label{table:dates}
		\begin{tabular}{ccccc}
			\hline
			\T Extremum  &   Start    &    End     & $\overline{F_{10.7}}$&\#GONG \\ 
			\B&&&$\left[\si{sfu}\right]$&datasets\\\hline
			\T Minimum 23 & 1995-05-07 & 1997-08-11& 73.7&23\\
			Maximum 23 & 2000-01-29 & 2002-05-05& 184.9&23\\
			Minimum 24 & 2007-09-21 & 2009-11-25& 70.0&22\\
			Maximum 24 & 2012-01-01 & 2014-04-30& 126.8&24\\
			\B Minimum 25 & 2018-01-01 & 2020-04-12& 70.4 & 19\\ \hline
		\end{tabular}
	\end{center}
\end{table}

\section{Data}\label{sec:2}
\subsection{GONG mode parameters}\label{sec:21}
At the time of writing, the publicly available GONG mode parameter files cover more than one full cycle of the solar magnetic field, i.e., more than two 11-year Schwabe cycles: the start date of the time series is May 7, 1995, the last available segment ends on April 12, 2020. We use the p-mode parameter tables, which were produced by the standard GONG pipeline from solar full-disk Dopplergrams \citep{Anderson1990, Hill1996, Hill1998} and are available online\footnote{\footurlone}. The parameters in these tables are obtained by fitting the power spectra of 108-day long GONG Doppler-velocity time series. Consecutive datasets overlap by 72 days, i.e., each third dataset is independent. Every 36 days a new dataset is added, which defines the unit of one GONG month. Thus, over one year about ten GONG months are accumulated.

In each GONG dataset's power spectrum, the oscillation multiplets are fitted with one symmetrical Lorentzian profile per azimuthal order $-l\leq m\leq l$ and a linear background for all harmonic degrees up to $l=150$. The free parameters of the fit model are mode frequency $\nu_{nlm}$, mode width $\Gamma_{nlm}$ (full width at half maximum), mode amplitude $A_{nlm}$, background offset $b_{0,nlm}$, and background slope $b_{1,nlm}$, where the triple $\left(nlm\right)$ indicates the radial order $n$, harmonic degree $l$, and azimuthal order $m$ of a mode. In this article, we concentrate on the product $\Gamma_{nlm}^2\cdot A_{nlm}$, which is, as we will see in further detail in Section~\ref{sec:32}, proportional to the mode energy supply rate $dE\slash dt_{nlm}$. 

\subsection{Solar radio flux $F_{10.7}$}\label{sec:22}
As a proxy for solar magnetic activity, we used the solar radio flux given by the $F_{10.7}$ index \citep{Tapping2013}, which is a good proxy for the level of activity in the upper chromosphere and the lower corona \citep{Tapping1987,Broomhall2015a}. Its time series is available online\footnote{\footurltwo}. For a measurement of $F_{10.7}$, the total emission on the solar disk at a wavelength of \SI{10.7}{\centi\metre} is integrated over one hour. It is given in solar flux units \si{sfu}, where \SI{1}{sfu} = \SI{e-22}{W.m^{-2}.Hz^{-1}}. The $F_{10.7}$ index is the averaged $F_{10.7}$ flux scaled for a distance of \SI{1}{AU}. 

There is no continuous time series that covers the entire time period that GONG has been in operation. Thus, we concatenate the tables \path{F107_1947_1996.txt}, \path{F107_1996_2007.txt}, and \path{fluxtable.txt} (data since 2004), which are available at the given URL, at the end date of each preceding table, ignoring the overlapping data. Measurements which are taken on the same day are averaged. These daily values were then averaged over the 108-day periods of the GONG datasets to yield $\overline{F_{10.7}}$.

In Figure~\ref{fig:f107}, the resulting $\overline{F_{10.7}}$ index is shown as a function of time, limited to the period covered by the available GONG data. The blue horizontal line indicates the median value of the $\overline{F_{10.7}}$ index through the used time frame, $\overline{F_{10.7}}_{\text{, median}}=\SI{95.6}{sfu}$. Data points above this value are coloured red, points below are black. We will use this colour code in later figures to ease the interpretation of the data. 

The start and end dates of the periods of solar activity minima and maxima through the GONG observations are listed in Table~\ref{table:dates}. These periods are shaded grey in Figure~\ref{fig:f107} and the respective identifier is printed at the top of the panel. Again, we will reuse this background colouring in later figures to ease identification of activity minima and maxima. The dates for the first three extrema are taken from \cite{Broomhall2017}. The dates of Max$_{24}$ were chosen such as to have a similar number of GONG datasets as the preceding maximum at times of high activity. The time when $\overline{F_{10.7}}$ has reached a comparable level of low activity as Min$_{24}$ was chosen as the starting date of the currently ongoing minimum. The number of GONG datasets during each of these periods are given in the last column of Table~\ref{table:dates}.


\section{Data Preparation}\label{sec:3}
The data processing steps and the corrections that were applied to the quantity $\Gamma_{nlm}^2\cdot A_{nlm}$ are the same as described in detail by \cite{Komm2000a, Komm2000}, and \cite{Kiefer2018a} for its constituent quantities $\Gamma_{nlm}$ and $A_{nlm}$. Here, we will briefly lay out the rationale of these corrections and refer to \cite{Kiefer2018a} for more details.

\subsection{Corrections and averaging}\label{sec:31}
\subsubsection*{Correction for spatial masking and azimuthal averaging}\label{sec:311}
Due to projection effects that increase towards the solar limb -- affecting pixel resolution as well as measured radial velocities -- mode amplitudes of azimuthal orders with lower $|m\slash l|$ are suppressed. This suppression follows an empirically determined polynomial in $\left(m\slash l\right)^{k}$ with $k=0,2,4$. To correct for this, we subtracted the fit from each multiplet and added the value of $|m\slash l|=1$. 

This fit, as all linear regressions in this article, was performed with \textsc{statsmodels}'s weighted linear regression routine \citep{seabold2010statsmodels}. In this particular case, the fit was done multilinearly in $\left(m\slash l\right)^{2}$ and $\left(m\slash l\right)^{4}$. The variance-weighted mean of the corrected data points is then adopted as the representative value of $\Gamma_{nl}^2\cdot A_{nl}$ of each multiplet. For a multiplet to be considered at all, a minimum of one third of azimuthal orders was required to be present for multiplets with $l>10$ and two thirds for multiplets with $l\leq10$. 

In the following, the index $m$ is dropped from all quantities, as the azimuthal orders of a multiplet have been averaged over. The frequency of the $m=0$ singlet is taken as the multiplet frequency. If the $m=0$ was not fitted, then the mean of the $m=\pm 1$ singlet frequencies is used. As the azimuthal orders have been averaged over, from here on out the phrase \textit{multiplet} will be avoided where possible and the term \textit{mode} will be used instead.

\subsubsection*{Correction for temporal window}\label{sec:312}
GONG is a network of six observing stations that are distributed around the Earth to maximise continuous viewing of the Sun. Still, the duty cycle (henceforth called fill) of the time series is lower than unity for all GONG datasets.\footnote{\footurlthree} The fill of the 108-day time series varies between a minimum of \SI{69.7}{\percent} and a maximum of \SI{93.7}{\percent} with a median value of \SI{86.7}{\percent}. This temporal window function artificially increases mode widths and decreases mode amplitudes. 

As the fill is rather high and the gaps are typically not ordered in a repeating pattern, it is sufficient to account for its impact by a linear regression \citep{Komm2000a}: $\Gamma_{nl}^2\cdot A_{nl}$ of each mode are fitted with a linear function as a function of fill, which is then subtracted and the extrapolated value at fill $=1$ is added.

\subsubsection*{Apparent solar radius}\label{sec:313}
Just as \cite{Kiefer2018a}, we corrected for residual annual changes of the solar angular radius. For this, we performed a linear fit of $\Gamma_{nl}^2\cdot A_{nl}$ as a function of the solar apparent radius and adjusted $\Gamma_{nl}^2\cdot A_{nl}$ to its extrapolated value at the maximum of the apparent solar radius during the time series.

\subsubsection*{Jumps in mode amplitudes}\label{sec:314}
\cite{Kiefer2018a} discovered two jumps in the mode amplitudes, which had to be corrected for with an empirical correction factor. The first jump occurs around the year 2001 and is due to an upgrade of the GONG hardware. The second jump, around 4 years later, is of uncertain origin. We continue using the correction factor of \cite{Kiefer2018a}, which they give in their appendix A, equation (12).

\begin{figure*}
	\centering
	\includegraphics[width=\linewidth]{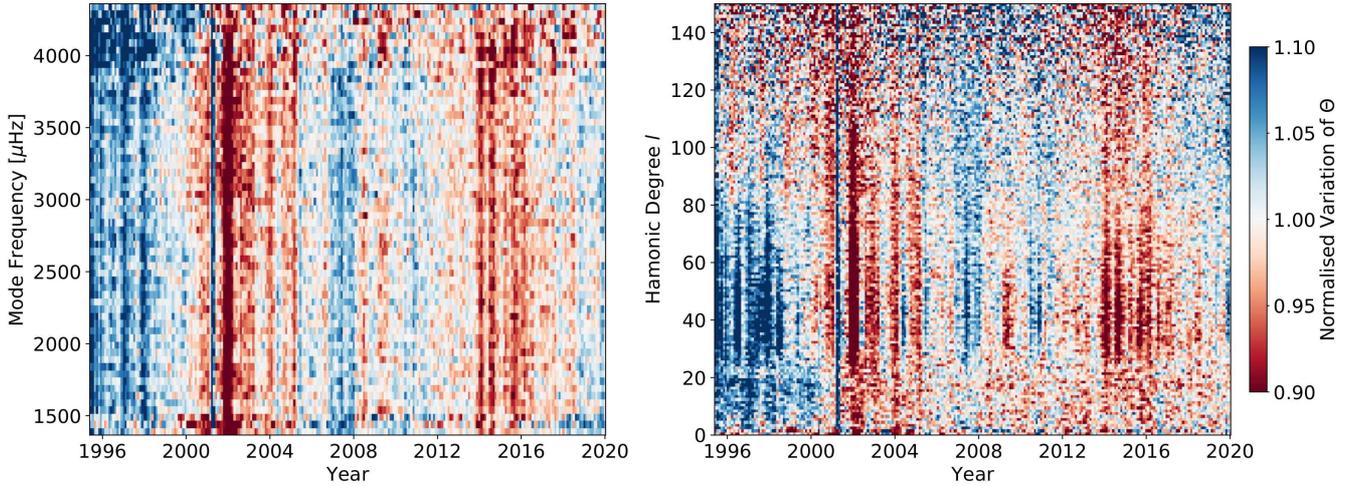}\caption{Variation of energy supply rate parameter $\Theta$ as a function of time and mode frequency (left panel) and of harmonic degree (right panel). $\Theta_{nl}$ of each mode is first normalised to the mean over the entire time series, then averages over independent bins of \SI{50}{\micro\hertz} (in the left panel) or individual harmonic degree (right panel) are calculated. The colour map is capped at $\left[0.9,1.1\right]$.}\label{fig:map}
\end{figure*}

\subsection{Energy supply rates}\label{sec:32}
Subsequently, we will consider either the energy supply rate parameter of individual modes, which, for brevity, we baptise $\Theta_{nl} = \Gamma_{nl}^2\cdot A_{nl}$, or the physical quantity of energy supply rate $dE\slash dt_{nl}$, see Eq.~(\ref{eq:dEdt}). If the average over modes in certain ranges of mode frequency and harmonic degrees is taken, we drop the indices $\left(nl\right)$.

\subsubsection*{Maps of the temporal variation of $\Theta$}\label{sec:321}
Figure~\ref{fig:map} shows the normalised variation of $\Theta$ as a function of time and mode frequency (left panel) and harmonic degree (right panel). Here, the modes' $\Theta_{nl}$ are first normalised to the mean over the entire time series. In the left panel, modes are grouped in independent bins of \SI{50}{\micro\hertz} and averaged. In the right panel, the average is taken for all modes of the same harmonic degree. The colour map is capped at $[0.9, 1.1]$. 

It can be seen that there is considerable temporal variation in $\Theta$, with bluer, i.e., larger values during activity minima as stated in Table~\ref{table:dates}. Red values, which indicate smaller $\Theta$, can predominantly be found at times of high magnetic activity. 

The rather sharp blue vertical feature at around the year 2001 is due to the GONG hardware upgrade at the time and can be considered an instrument artefact. 

From these two panels, it can already be estimated that any cyclic variation in the mode energy supply rates is likely less apparent in modes with frequencies $\gtrsim\SI{3500}{\micro\hertz}$ and harmonic degrees $\gtrsim~120$. Also, even though the variation with apparent solar radius has been accounted for, a seemingly repeating, quasi-yearly pattern can be identified. We will discuss this further in Section~\ref{sec:6}.

\begin{figure}
	\centering
	\includegraphics[width=\linewidth]{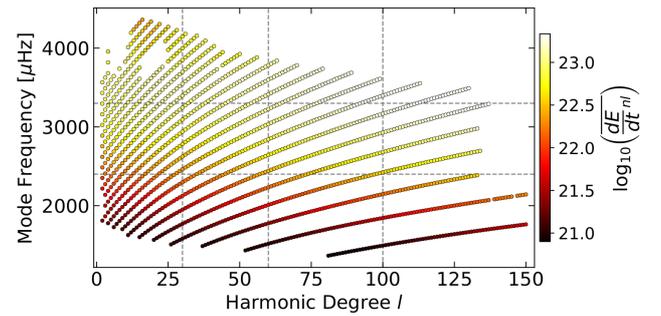}\caption{$l$-$\nu$-diagram of modes which are present in at least \SI{50}{\percent} of all time samples. The colour map gives the decadic logarithm of the time averaged energy supply rate. Mode inertiae are normalised at the GONG observation height $r_{\text{obs}}$. The vertical and horizontal dashed grey lines indicate the boundaries between mode sets which we use in later sections.
	}\label{fig:erate_ellnu}
\end{figure}

\subsubsection*{Conversion to energy per unit time}\label{sec:322}
From the measured quantity $\Theta_{nl}=\Gamma_{nl}^2\cdot A_{nl}$, the energy that is supplied to the p-mode oscillations per mode can be calculated by \citep[see, e.g.,][]{Goldreich1994}
\begin{align}
\left(\frac{dE}{dt}\right)_{nl} = \pi^2 C_{\text{vis}}M_{nl}\Gamma_{nl}^2 A_{nl}, \label{eq:dEdt}
\end{align}
where $\Gamma_{nl}$ is in units of \si{\hertz}, $A_{nl}$ has units \si{\centi\metre^{2}\second^{-2}\hertz^{-1}}. The numerical factor $C_{\text{vis}}=3.33$ corrects for the GONG specific reduced visibility of modes due to leakage effects \citep{Hill1998}. We will comment on the use of this factor in Section~\ref{sec:6.3}. The factor $\pi^2$ stems from the Lorentzian profile that is used to fit the mode peaks and the assumption of scattering being the main contribution to mode damping. The mass $M_{nl}$ of the mode $\left(nl\right)$ is calculated as 
\begin{align}
M_{nl} = 4\pi M_{\odot} I_{nl},\label{eq:modemass}
\end{align}
with the solar mass $M_{\odot}$ in \si{\gram} and the mode inertia of solar Model S \citep{Christensen-Dalsgaard1996}, which is defined as
\begin{align}
I_{nl} = \frac{\int_{0}^{R_{\odot}}\rho\left(\left|\xi^r_{nl}\right|^2+l\left(l+1\right)\left|\xi^h_{nl}\right|^2\right)r^2dr}{M_{\odot}\left(\left|\xi^r_{nl}\left(r'\right)\right|^2+l\left(l+1\right)\left|\xi^h_{nl}\left(r'\right)\right|^2\right)},\label{eq:inertia}
\end{align}
where $\rho$ is density, $\xi^r_{nl}$ and $\xi^h_{nl}$ are the mode's radial and horizontal displacement eigenfunctions, $R_{\odot}$ is the solar radius, $r$ is the radial coordinate, and $r'$ is the radial position at which the normalisation is done. We normalise the mode inertiae at either the photospheric radius $r'=R_{\odot}$ or the GONG observation height at $r'=R_{\odot}+\SI{240}{km}\equiv r_{\text{obs}}$ \citep{Baudin2005}. Unless stated otherwise, mode inertiae are normalised at the GONG observation height $r_{\text{obs}}$. Values normalised at $R_{\odot}$ are provided in an online-supplement table.

We use \si{cgs} units, i.e., $dE\slash dt_{nl}$ has units \si{erg.s^{-1}}. More detailed discussions of how Eq.~(\ref{eq:dEdt}) comes about can be found in, e.g., \cite{Goldreich1994} and \cite{Komm2000}. We note that normalised variations of $dE\slash dt_{nl}$ as a function of time are equivalent to normalised variations of $\Theta_{nl}$ in time, as the factors $\pi^2 C_{\text{vis}}M_{nl}$ then cancel.

Figure~\ref{fig:erate_ellnu} shows an $l$-$\nu$-diagram of the 1580 modes which are present in at least \SI{50}{\percent} of the GONG datasets. We excluded modes with $l=0,1$ as they appear as outliers in the distribution of supply rates (see Figure~\ref{fig:erate_avg}). There are 26 radial and dipole modes, i.e., the total number of modes with a presence rate of at least \SI{50}{\percent} is 1606. The colour of the points indicates $\log_{10}\left(\overline{dE\slash dt_{nl}}\right)$, where the variance-weighted average over all time samples (indicated by the overline) is taken as the representative value for each mode. The vertical and horizontal dashed grey lines indicate the boundaries in mode frequencies and harmonic degrees that separate mode sets which we use in later sections.

The top left panel of Figure~\ref{fig:erate_avg} shows the energy supply rates of all modes as a function of mode frequency. Again, the variance-weighted average over all time samples is taken. Squares highlight radial modes, triangles are used to accent dipole modes. Here it can clearly be seen that radial and dipole modes are outliers from the general bulk of data. This problem arises from the non-inclusion of the instrument-specific leakage matrix in the GONG fitting pipeline. We do not adopt individual correction factors for different harmonic degrees in this article (consistently with \citealt{Komm2000a}). From here on out we shall not consider radial and dipole modes any more.

In the bulk of the data points, ridges can be identified. These are made up of modes of like radial order $n$, as is highlighted by the discrete colour map. These ridges can be seen more clearly when $\overline{dE\slash dt_{nl}}$ is plotted as a function of harmonic degree $l$, see the bottom left panel of Figure~\ref{fig:erate_avg}.

In Table~\ref{table:erates} we give the energy supply rates for averages over 32 modes consecutive in mode frequency.  The second and third columns of this table are energy supply rates without and with rescaling by $Q$ averaged over the entire time series. From the second column we find that energy supply rates peak at around \SI{3460}{\micro\hertz} with a value of \SI{13.394(4)e22}{erg.s^{-1}}. The entries in the following columns are averaged over periods of activity extrema, i.e., all activity minima or maxima as defined in Table~\ref{table:dates}, and scaled by $Q$ in the last two columns. This table is also available in machine-readable form online.

\subsubsection*{Accounting for different mode inertiae}\label{sec:323}
As was shown by, e.g., \cite{Komm2002}, when plotted as a function of mode frequency the ridges of individual radial orders are collapsed almost completely into one ridge by multiplication of $\overline{dE\slash dt_{nl}}$ with the scaling factor
\begin{align}
Q_{nl} = \frac{I_{nl}}{\overline{I_{n0}}\left(\nu_{nl}\right)}. \label{eq:Q}
\end{align}
Here, the radial mode inertia $I_{n0}$ is interpolated to the mode frequency $\nu_{nl}$ to obtain $\overline{I_{n0}}(\nu_{nl})$.

\begin{figure*}
	\centering
	\includegraphics[width=\linewidth]{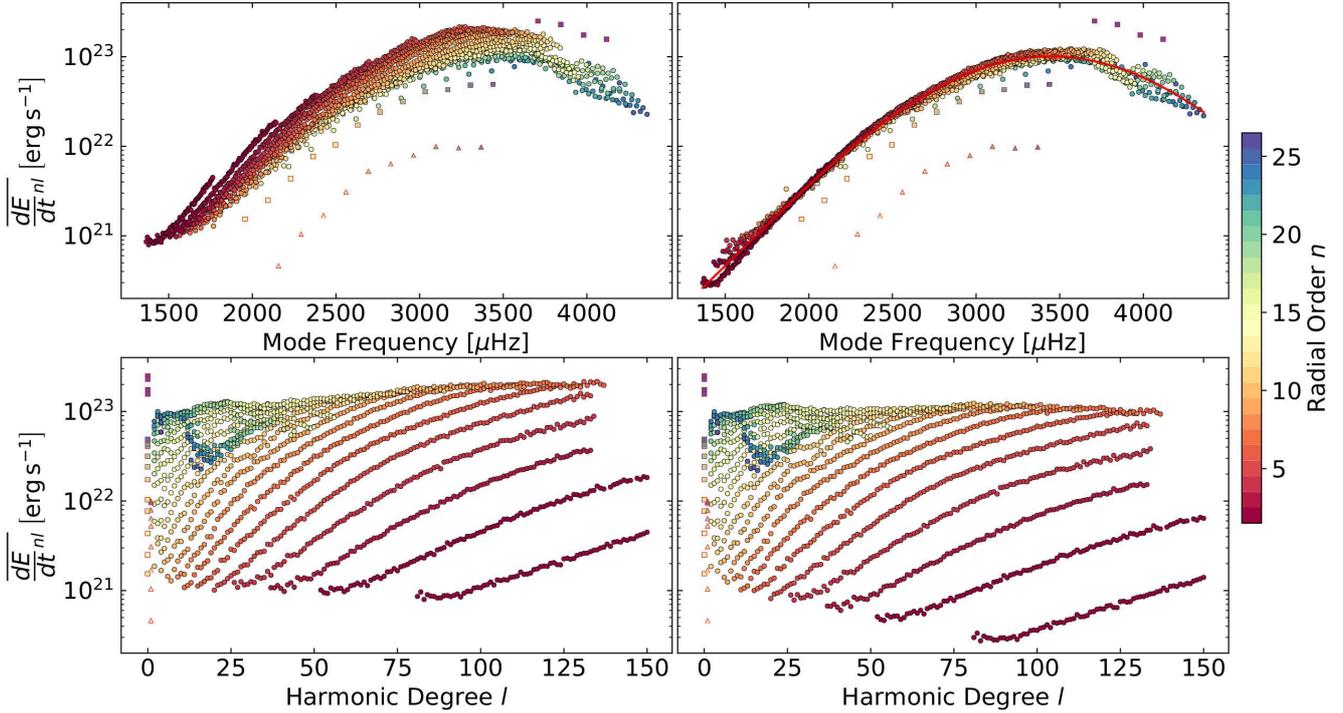}
	\caption{Left column: Energy supply rates calculated with Eq.~(\ref{eq:dEdt}) as function of mode frequency (top panel) and harmonic degree (bottom panel). Right column: As left column, but supply rates are inertia-scaled by multiplication with Eq.~(\ref{eq:Q}). Top right panel: The solid red line is a fit of the exponential function Eq.~(\ref{eq:gauss}) to the supply rates. All panels: Radial orders are indicated by the colours of the data points. Radial and dipole modes are highlighted by red-bordered triangles and squares, respectively.}\label{fig:erate_avg}
\end{figure*}

The right column of panels in Figure~\ref{fig:erate_avg} shows the time-averaged energy supply rate values as in the left column, but scaled by multiplication of $Q_{nl}$. Now, as can be seen in the top right panel of Figure~\ref{fig:erate_avg}, the supply rates can well be described by a function which depends only on mode frequency $\nu$. We used a non-linear least squares optimization to fit the exponential
\begin{align}
G\left(\nu, p\right) = a + b\cdot\exp\left(-\frac{\left(\nu-\nu_0\right)^2}{2\gamma^2}\right) \label{eq:gauss}
\end{align}
to the base-10 logarithm of the resulting energy supply rates. This fit is shown by the solid red line. In Eq.~(\ref{eq:gauss}), the parameter tuple $p$ comprises: offset $a$, magnitude $b$, central frequency $\nu_0$, and width of the exponential $\gamma$. The best-fit parameters and their standard uncertainties are given in Table~\ref{table:gauss}. Though present in the panel, the radial and dipole modes are excluded from the fit. 

We also tested an asymmetric and a symmetric Voigt profile, both of which had a comparable goodness-of-fit as the exponential of Eq.~(\ref{eq:gauss}). Larger deviations from the exponential behaviour occur at low mode frequencies \mbox{$\lesssim\SI{1800}{\micro\hertz}$} and at high mode frequencies \mbox{$\gtrsim\SI{3600}{\micro\hertz}$}. At low frequencies, this is at least partially caused by the limited frequency resolution of the 108-day time series which is $\approx\SI{0.107}{\micro\hertz}$. Thus, only a few frequency bins are available per mode width which become smaller with decreasing mode frequency, see, e.g., \cite{Komm2000} and \cite{Kiefer2018a}. This can lead to an imprecise estimation of mode widths and thus also of $\overline{dE\slash dt_{nl}}$. At high frequencies $\gtrsim\SI{3600}{\micro\hertz}$ some radial orders depart from the monotonic increase with mode frequency. This can be appreciated in the bottom panels of Fig.~\ref{fig:erate_avg}, where the data points with $n\gtrsim 20$ show a dip in $\overline{dE\slash dt_{nl}}$.

In a machine-readable online-only table, we provide the energy supply rates of all 1606 modes which are present in at least \SI{50}{\percent} of the GONG samples. There, we provide mode inertiae and scaling factors $Q$ normalised at both the photospheric radius $R_{\odot}$ and the GONG observation height in the solar atmosphere $r_{\text{obs}}$. The energy supply rates are then provided for 12 different averages: averaged over all time samples with and without scaling by $Q$, averaged over times of the activity minima and maxima, both with and without $Q$-scaling, and all of these at both normalisation heights $R_{\odot}$ and $r_{\text{obs}}$.

\begin{table}
	\caption{Fit parameters and uncertainties of exponential function Eq.~(\ref{eq:gauss}) to the inertia corrected energy supply rates.}\label{table:gauss}
	\begin{center}
		\begin{tabular}{ccccc}
			\hline
			\T&$a$& $b$ & $\nu_0$  & $\gamma$ \\
			\B&&&$\left[\si{\micro\hertz}\right]$&$\left[\si{\micro\hertz}\right]$\\
			\hline
			\T Value & 17.22 & 5.79  & 3446.7 & 1913.1 \\
			\B Uncertainty & 0.14	& 0.14 & 3.1 & 29.3 \\
			\hline
		\end{tabular}
	\end{center}
\end{table}


\section{Results}\label{sec:4}
\subsection{Averages over parameter ranges}\label{sec:41}
As can already be appreciated from Figure~\ref{fig:map}, there is a temporal variation in the energy supply rates. In order to decrease the uncertainties, we take variance-weighted averages over four mode frequency ranges and five ranges in harmonic degree. These ranges are given in the first four columns of Table~\ref{table:amp} and are indicated by grey dashed lines in Figure~\ref{fig:erate_ellnu}. 

As we set the presence rate of modes to \SI{50}{\percent}, not all modes are present for all time samples. The range of number of modes within each parameter range that is averaged over is given in the fifth column of Table~\ref{table:amp}. 

To select the modes which contribute to the variance-weighted average within a parameter range we do the following: First, we select all the modes, which are within the $\nu_{\text{min}}$-$\nu_{\text{max}}$ and ${l_{\text{min}}\text{-}l_{\text{max}}}$ ranges. The energy supply rates of each mode are then normalised to the variance-weighted mean over the entire time series. Modes which do not satisfy the \SI{50}{\percent} quota are eliminated from the set. At each time sample we then remove all modes whose normalised variation is not within the interval $\left[0.5, 1.5\right]$, as very large deviations from 1 are not expected to occur. Further, we remove all modes that deviate by more than four times the standard deviation of the contributing modes at each time sample. These two outlier rejections only remove a handful of modes per time sample, if any.

\begin{figure*}
	\centering
	\includegraphics[width=\linewidth]{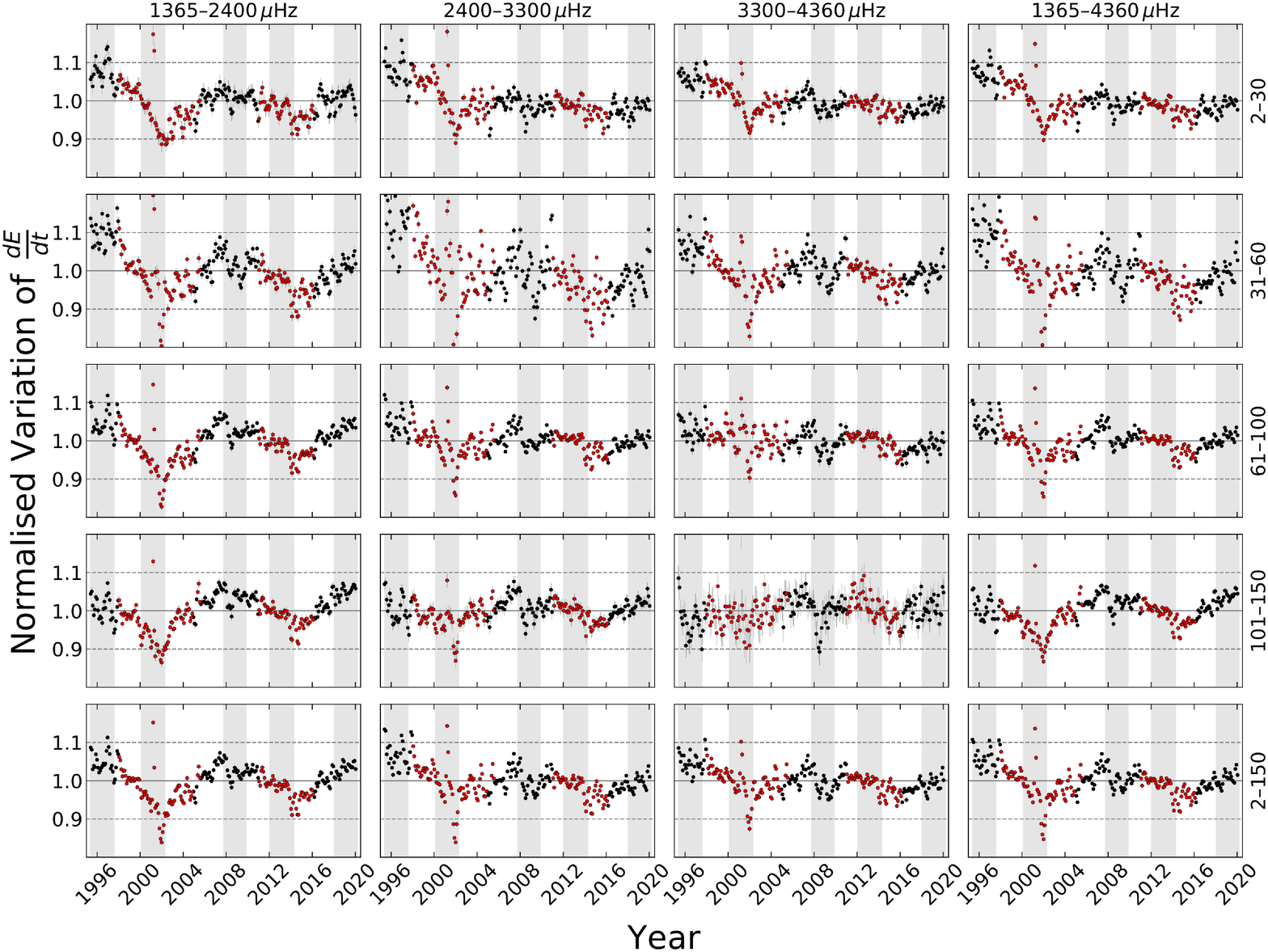}\caption{Temporal variation of energy supply rates for different ranges of harmonic degrees (rows; harmonic degree range indicated to the right of the fourth column) and mode frequencies (columns; frequency range indicated above the first row). Mode energy supply rates are normalised to the mean for each mode and then the variance-weighted average over all modes in the respective range of frequency and degree is taken. Data with higher than median $\overline{F_{10.7}}$ solar radio flux are highlighted by red points. Levels of 1.1 and 0.9 of the mean are indicated by dashed lines. Grey shaded times corresponds to the activity minima and maxima laid out in Figure~\ref{fig:f107} and Table~\ref{table:dates}.}\label{fig:matrix}
\end{figure*}

The variance-weighted averages of the normalised variation of $dE\slash dt$ over these mode sets are presented in Figure~\ref{fig:matrix}. Parameter ranges of the mode set entering each panel are indicated to the top of each column and the right of each row. The error bars indicate $5\sigma$, i.e., uncertainties on the individual data points are very small due to the averaging over azimuthal orders and the modes in each range.

The minimum and maximum variations of each panel are given in the sixth and seventh column of Table~\ref{table:amp}. To reduce short-term variation caused by seasonal variations in data quality that are not captured by the fill factor correction, these values are calculated from the one-year boxcar smoothed data. The median uncertainty of the unsmoothed data is given in the last column of Table~\ref{table:amp}. The temporal variation of $dE\slash dt$ is statistically significant in all 20 parameter ranges. 

\begin{table*}
	\centering
	\caption{Normalised and averaged variation of energy supply rates for modes sets with different parameter ranges (defined in columns 1--4) that are presented in the panels of Fig.~\ref{fig:matrix}. Column 5 gives the range of number of modes included in the GONG datasets. Columns 6 and 7 give the extrema of the one-year smoothed variation and their uncertainty. The last column gives the median error of the individual un-smoothed energy supply rate in this parameter range.}\label{table:amp}
	\begin{tabular}{cccccccc}
		\hline
		\multicolumn{2}{c}{Frequency range [\si{\micro\hertz}]} & \multicolumn{2}{c}{Harmonic degrees}  &    Number of     & \multicolumn{2}{c}{Normalised amplitude [\%]} &  Median error \T\\ \cline{1-2}\cline{3-4}\cline{6-7}
		$\nu_{\text{min}}$                                         & $\nu_{\text{max}}$ & $l_{\text{min}}$ & $l_{\text{max}}$ &   modes    &               min                &      max       & [\%] \B \\ \hline
		\T1365 &  2400 & 2 & 30 & 94--150 & 90.0$\pm$0.1 & 109.3$\pm$0.1 & 0.4 \\
		2400 &  3300 & 2 & 30 & 166--185 & 92.9$\pm$0.1 & 109.8$\pm$0.1 & 0.2  \\
		3300 &  4360 & 2 & 30 & 125--150 & 94.1$\pm$0.1 & 107.1$\pm$0.1 & 0.3  \\
		\B1365 &  4360 & 2 & 30 & 412--482 & 92.8$\pm$0.1 & 108.4$\pm$0.1 & 0.2  \\ \hline
		\T1365 &  2400 & 31 & 60 & 101--137 & 89.2$\pm$0.1 & 110.5$\pm$0.1 & 0.3   \\
		2400 &  3300 & 31 & 60 & 94--152 & 89.2$\pm$0.1 & 116.6$\pm$0.1 & 0.2   \\
		3300 &  4360 & 31 & 60 & 119--128 & 90.1$\pm$0.1 & 107.4$\pm$0.1 & 0.2   \\
		\B1365 &  4360 & 31 & 60 & 357--417 & 89.8$\pm$0.1 & 111.4$\pm$0.1 & 0.1   \\ \hline
		\T1365 &  2400 & 61 & 100 & 126--150 & 87.5$\pm$0.1 & 105.9$\pm$0.1 & 0.2   \\
		2400 &  3300 & 61 & 100 & 147--162 & 92.2$\pm$0.1 & 107.6$\pm$0.1 & 0.1   \\
		3300 &  4360 & 61 & 100 & 81--88 & 95.5$\pm$0.1 & 105.5$\pm$0.1 & 0.2   \\
		\B1365 &  4360 & 61 & 100 & 369--400 & 91.5$\pm$0.1 & 106.7$\pm$0.1 & 0.1   \\ \hline
		\T1365 &  2400 & 101 & 150 & 122--131 & 88.9$\pm$0.1 & 106.1$\pm$0.1 & 0.2    \\
		2400 &  3300 & 101 & 150 & 38--115 & 93.1$\pm$0.1 & 105.2$\pm$0.1 & 0.3   \\
		3300 &  4360 & 101 & 150 & 16--32 & 95.1$\pm$0.2 & 104.0$\pm$0.2 & 0.6   \\
		\B1365 &  4360 & 101 & 150 & 183--278 & 90.4$\pm$0.1 & 105.5$\pm$0.1 & 0.2    \\ \hline
		\T1365 &  2400 & 2 & 150 & 443--566 & 88.6$\pm$0.1 & 105.9$\pm$0.1 & 0.1    \\
		2400 &  3300 & 2 & 150 & 501--614 & 91.9$\pm$0.1 & 109.5$\pm$0.1 & 0.1    \\
		3300 &  4360 & 2 & 150 & 371--398 & 93.5$\pm$0.1 & 106.3$\pm$0.1 & 0.1   \\
		\B1365 &  4360 & 2 & 150 & 1381--1570 & 91.2$\pm$0.1 & 107.6$\pm$0.1 & 0.1    \\ \hline
	\end{tabular}
\end{table*}

\subsection{Low harmonic degrees}\label{sec:42}
\begin{figure}
	\centering
	\includegraphics[width=\linewidth]{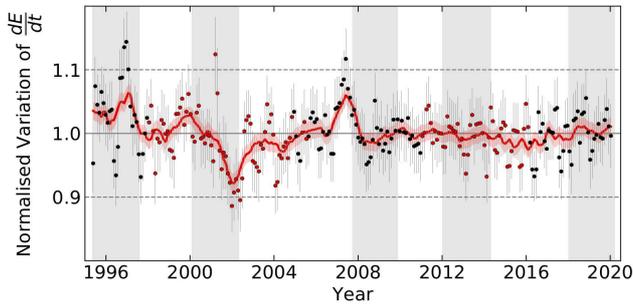}\caption{Normalised variation of the energy supply rate of modes of harmonic degrees $l=2\text{--}4$ as a function of time. Colour coding as in Figure~\ref{fig:matrix}.}\label{fig:lowl}
\end{figure}

To emulate the observations of disk-integrated helioseismic observations, we also averaged the lowest available harmonic degrees $l=2\text{--}4$ over the complete frequency range. Figure~\ref{fig:lowl} shows the resulting normalised variation as a function of time. Due to the smaller number of azimuthal orders per multiplet and the smaller number of multiplets overall (between 28 and 43, depending on the time sample), the averages are less well constrained than in the wider harmonic degree ranges shown in Figure~\ref{fig:matrix}. The solid red line shows the running one-year variance-weighted mean with its uncertainty in the red shaded band. As in Fig.~\ref{fig:matrix} the error bars indicate $5\sigma$ uncertainties.

The extrema of these smoothed data are \SI{92.2(3)}{\percent} and \SI{106.4(3)}{\percent}, where the median uncertainty on the un-smoothed data point is \SI{0.9}{\percent}. We will discuss this further in Section~\ref{sec:6}.

\subsection{Cross-correlating magnetic activity and $\frac{dE}{dt}$}\label{sec:43}
\begin{figure*}
	\centering
	\includegraphics[width=\linewidth]{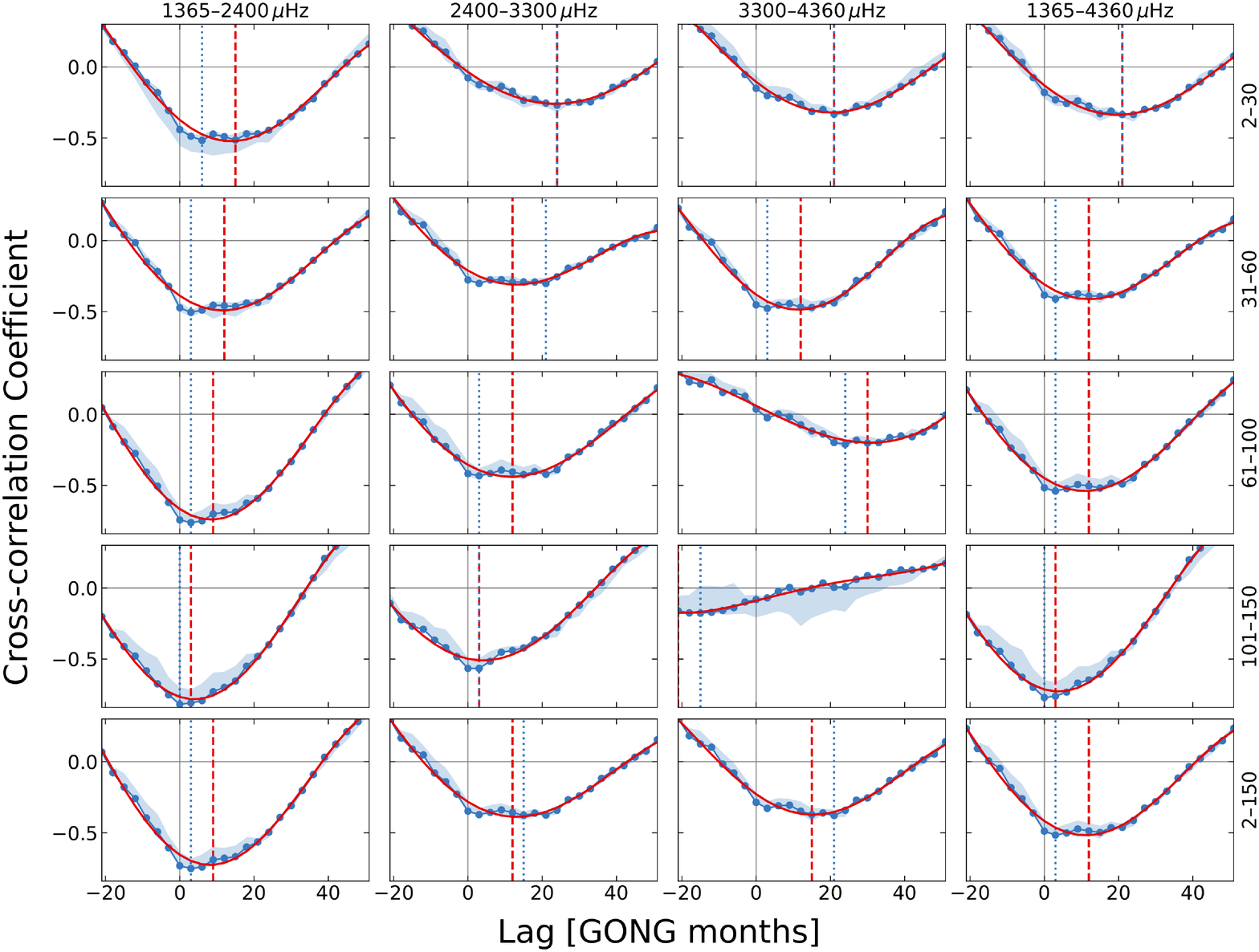}\caption{Cross-correlation functions (CCF) between the data in the corresponding panels of Fig.~\ref{fig:matrix} and the $\overline{F_{10.7}}$ solar radio flux of Fig.~\ref{fig:f107} (blue curves and blue circles). The minimum value of each panel's CCF is indicated by a vertical blue dotted line. Red curves show fourth-order polynomial fits to the CCF in the presented range of lags. The minimum value of each panel's fit is indicated by a vertical red dashed line. The ordinate extends from -0.84 to 0.3 in all panels. The abscissa, i.e., the lag, is in units of GONG months, where one GONG month equals \unit[36]\si{\day}. The blue shaded bands show the minimum-to-maximum range of the three independent sets of data of which the first set was chosen for the fit.}\label{fig:ccf} 
\end{figure*}

To assess the correlation between the variation observed in $dE\slash dt$ and the level of solar magnetic activity, as measured by the $\overline{F_{10.7}}$ index, we computed the cross-correlation function (CCF) for the independent time samples, i.e., every third time step. We again split the modes into the subsets we used before: the modes included in the calculation of the functions shown in the panels of Figure~\ref{fig:ccf} correspond to those of Figure~\ref{fig:matrix}. In Figure~\ref{fig:ccf}, the cross-correlation functions are shown by solid blue lines with blue dots. The first of the three possible sets of independent data points was chosen for analysis.  Fourth-order polynomial fits to the CCFs are shown by the solid red curves. The dotted vertical blue line indicates the global minimum of each panel's CCF, while the dashed vertical red line indicates the minimum of the fitted function. These fits are performed with \textsc{numpy}'s least-square polynomial fit routine \textit{numpy.polyfit} \citep{Oliphant2006}. The blue shaded bands indicate the minimum-to-maximum range of the three possible independent data sets. The correlation values typically vary only by a few percent between the three data sets.

In Table~\ref{table:ccf} the Pearson correlation coefficient $r$ and its $p$-value for the different mode sets are given for different lag values: for un-lagged data in columns 5 and 6; for the lag of the minimum of the cross-correlation function, in columns 7--9; for the lag of the minimum of the quadratic fit to the cross-correlation function, in columns 10--12. The last two columns give Spearman's rank correlation $\rho$ for un-lagged data and its corresponding $p$-value.

For the un-lagged data the $p$-values are $<0.05$ for 15 out of the 20 mode sets. For low harmonic degrees $l=2\text{--}30$ only the low frequency modes have $p<0.05$. In addition, for the high frequency modes, the ranges $l=61\text{--}100$ and $l=101\text{--}150$ have $p>0.05$.

This essentially also holds true for the other lag values (global minimum of the CCF and minimum of the fitted function) and the Spearman $\rho$: For the low-, and mid-frequency sets, as well as for the full mode set, the $p$-value of the correlation coefficients is $<0.05$ for all harmonic degree ranges. Indeed, overall the correlation coefficients indicate a moderate to strong anti-correlation between the level of solar magnetic activity and the p-mode energy supply rates. 

The strongest anti-correlation with $r=-0.82$ is found for the low frequency mode set with $l=101\text{--}150$ at zero lag. Except for the high harmonic degree sets, the minima of all (statistically significant) CCFs are found at positive lag values.

\begin{table*}
	\caption{Correlation coefficients between mode energy supply rates of different parameter ranges (defined in columns 1--4) and the $\overline{F_{10.7}}$ solar radio flux for different lags: Pearson $r$ and corresponding $p$-value for un-lagged data in columns 5 and 6. Lag of the minimum of the cross-correlation function, this lag's Pearson $r$ and $p$-value in columns 7--9. The lag of the minimum of the polynomial fit to the cross-correlation function, this lag's Pearson $r$ and $p$-value in columns 10--12. Spearman's rank correlation $\rho$ for un-lagged data and corresponding $p$-value in columns 13 and 14. }\label{table:ccf}
	\centering
	\begin{tabular}{cccccccccccccc}
		\hline
		\multicolumn{2}{c}{Frequency range [\si{\micro\hertz}]} & \multicolumn{2}{c}{Harmonic degrees}  &      $r(0)$      & $p(0)$ & Lag$_{\text{min}}$ [\si{\day}] & $r_{\text{min}}$ & $p_{\text{min}}$ & Lag$_{\text{fit}}$ [\si{\day}] & $r_{\text{fit}}$ & $p_{\text{fit}}$ &   $\rho$   &   p  \T\\ \cline{1-2}\cline{3-4}
		$\nu_{\text{min}}$                                              & $\nu_{\text{max}}$ & $l_{\text{min}}$ & $l_{\text{max}}$ &        &                        &                  &                  &                        &                  &                  &            &       &            \B\\ \hline
		\T 1365 &  2400 & 2 & 30 & -0.43 & $<10^{-3}$ &216 & -0.51 & $<10^{-3}$ & 540 & -0.51 & $<10^{-3}$ & -0.44 & $<10^{-3}$ \\
		2400 &  3300 & 2 & 30 & -0.08 & 0.49 &864 & -0.27 & 0.019 & 864 & -0.27 & 0.019 & -0.02 & 0.84 \\
		3300 &  4360 & 2 & 30 & -0.15 & 0.17 &756 & -0.33 & 0.0029 & 756 & -0.33 & 0.0029 & -0.07 & 0.51 \\
		\B 1365 &  4360 & 2 & 30 & -0.18 & 0.1 &756 & -0.34 & 0.0029 & 756 & -0.34 & 0.0029 & -0.12 & 0.27 \\\hline
		\T 1365 &  2400 & 31 & 60 & -0.48 & $<10^{-3}$ &108 & -0.51 & $<10^{-3}$ & 432 & -0.47 & $<10^{-3}$ & -0.50 & $<10^{-3}$ \\
		2400 &  3300 & 31 & 60 & -0.29 & 0.0082 &756 & -0.31 & 0.0053 & 432 & -0.30 & 0.006 & -0.24 & 0.025 \\
		3300 &  4360 & 31 & 60 & -0.46 & $<10^{-3}$ &108 & -0.48 & $<10^{-3}$ & 432 & -0.48 & $<10^{-3}$ & -0.37 & $<10^{-3}$ \\
		\B 1365 &  4360 & 31 & 60 & -0.39 & $<10^{-3}$ &108 & -0.42 & $<10^{-3}$ & 432 & -0.40 & $<10^{-3}$ & -0.34 & 0.0013 \\\hline
		\T 1365 &  2400 & 61 & 100 & -0.75 & $<10^{-3}$ &108 & -0.76 & $<10^{-3}$ & 324 & -0.71 & $<10^{-3}$ & -0.73 & $<10^{-3}$ \\
		2400 &  3300 & 61 & 100 & -0.42 & $<10^{-3}$ &108 & -0.44 & $<10^{-3}$ & 432 & -0.41 & $<10^{-3}$ & -0.31 & 0.0037 \\
		3300 &  4360 & 61 & 100 & 0.03 & 0.8 &864 & -0.22 & 0.061 & 1080 & -0.21 & 0.072 & 0.06 & 0.61 \\
		\B 1365 &  4360 & 61 & 100 & -0.52 & $<10^{-3}$ &108 & -0.54 & $<10^{-3}$ & 432 & -0.51 & $<10^{-3}$ & -0.44 & $<10^{-3}$ \\\hline
		\T 1365 &  2400 & 101 & 150 & -0.82 & $<10^{-3}$ &0 & -0.82 & $<10^{-3}$ & 108 & -0.81 & $<10^{-3}$ & -0.79 & $<10^{-3}$ \\
		2400 &  3300 & 101 & 150 & -0.57 & $<10^{-3}$ &0 & -0.57 & $<10^{-3}$ & 108 & -0.57 & $<10^{-3}$ & -0.48 & $<10^{-3}$ \\
		3300 &  4360 & 101 & 150 & -0.10 & 0.36 &-540 & -0.18 & 0.1 & -756 & -0.17 & 0.1 & -0.05 & 0.67 \\
		\B 1365 &  4360 & 101 & 150 & -0.77 & $<10^{-3}$ &0 & -0.77 & $<10^{-3}$ & 108 & -0.76 & $<10^{-3}$ & -0.70 & $<10^{-3}$ \\\hline
		\T 1365 &  2400 & 2 & 150 & -0.74 & $<10^{-3}$ &108 & -0.75 & $<10^{-3}$ & 324 & -0.69 & $<10^{-3}$ & -0.72 & $<10^{-3}$ \\
		2400 &  3300 & 2 & 150 & -0.35 & $<10^{-3}$ &540 & -0.38 & $<10^{-3}$ & 432 & -0.37 & $<10^{-3}$ & -0.25 & 0.022 \\
		3300 &  4360 & 2 & 150 & -0.30 & 0.0064 &756 & -0.39 & $<10^{-3}$ & 540 & -0.38 & $<10^{-3}$ & -0.20 & 0.072 \\
		\B 1365 &  4360 & 2 & 150 & -0.50 & $<10^{-3}$ &108 & -0.52 & $<10^{-3}$ & 432 & -0.50 & $<10^{-3}$ & -0.43 & $<10^{-3}$ \\\hline
	\end{tabular}
\end{table*}

\subsection{Comparison of activity extrema}\label{sec:44}
We compare the energy supply rates during the different activity extrema listed in Table~\ref{table:dates} with each other. For this, we average the energy supply rates of each mode over the time samples during the periods of Table~\ref{table:dates}, take the difference between two extremal periods, and normalise by the mean over the entire time series. As activity and energy supply rates are anti-correlated, we subtract successive periods of high activity from periods of low activity for better comparability of the results. Also, to assess the relative depth or strength of minima and maxima, we compare the minima of cycles 25 and 24 as well as the maxima of cycles 24 and 23. Only energy supply rates of modes which are present in two periods can be subtracted from each other. Thus, if two rows of Table~\ref{table:extr_change_freq_org} are subtracted from each other, e.g. row three from row two to seemingly yield Max$_{24}$-Max$_{23}$, there is a small difference to the values given in the last row. In this example, the last row only necessitates the modes to be present in Max$_{24}$ and Max$_{23}$, whereas taking the difference between rows three and two demands the modes to be present in all of these rows' four extrema.

In Figure~\ref{fig:fracnu_unlagged}, the normalised differences between the different extrema are shown for all modes as a function of mode frequency (black data points). The extrema that are compared in the respective panel are indicated to the left of each panel. In order to smooth out the scatter in the normalised differences, we calculated the rolling variance-weighted average over 100 modes consecutive in mode frequency. In each panel, this is shown by the solid coloured curve and its $10\sigma$ confidence interval is given by the coloured band. The $1\sigma$ uncertainty is calculated as the standard error of the weighted mean. Due to the heavy averaging that has been applied up to this point, the uncertainties are rather small. Figure~\ref{fig:fracell_unlagged} shows the same as Figure~\ref{fig:fracnu_unlagged} but as a function of harmonic degree. Here, the weighted averaging is first done over the modes of each individual degree and then the rolling variance-weighted average over five consecutive degrees is calculated.

For better comparability, Figure~\ref{fig:diff_fraction} collects the smoothed curves from Figures~\ref{fig:fracnu_unlagged} and \ref{fig:fracell_unlagged}. Here, the left panel shows the normalised differences as a function of mode frequency and the right panel as a function of harmonic degree.

In Tables~\ref{table:extr_change_freq_org} and \ref{table:extr_change_harmdeg_org}, the percental energy supply rate change of modes in the four mode frequency ranges and five harmonic degree ranges are listed. We will discuss these results further in the next section.

\afterpage{
\begin{figure*}
	\centering
	\includegraphics[width=0.49\linewidth]{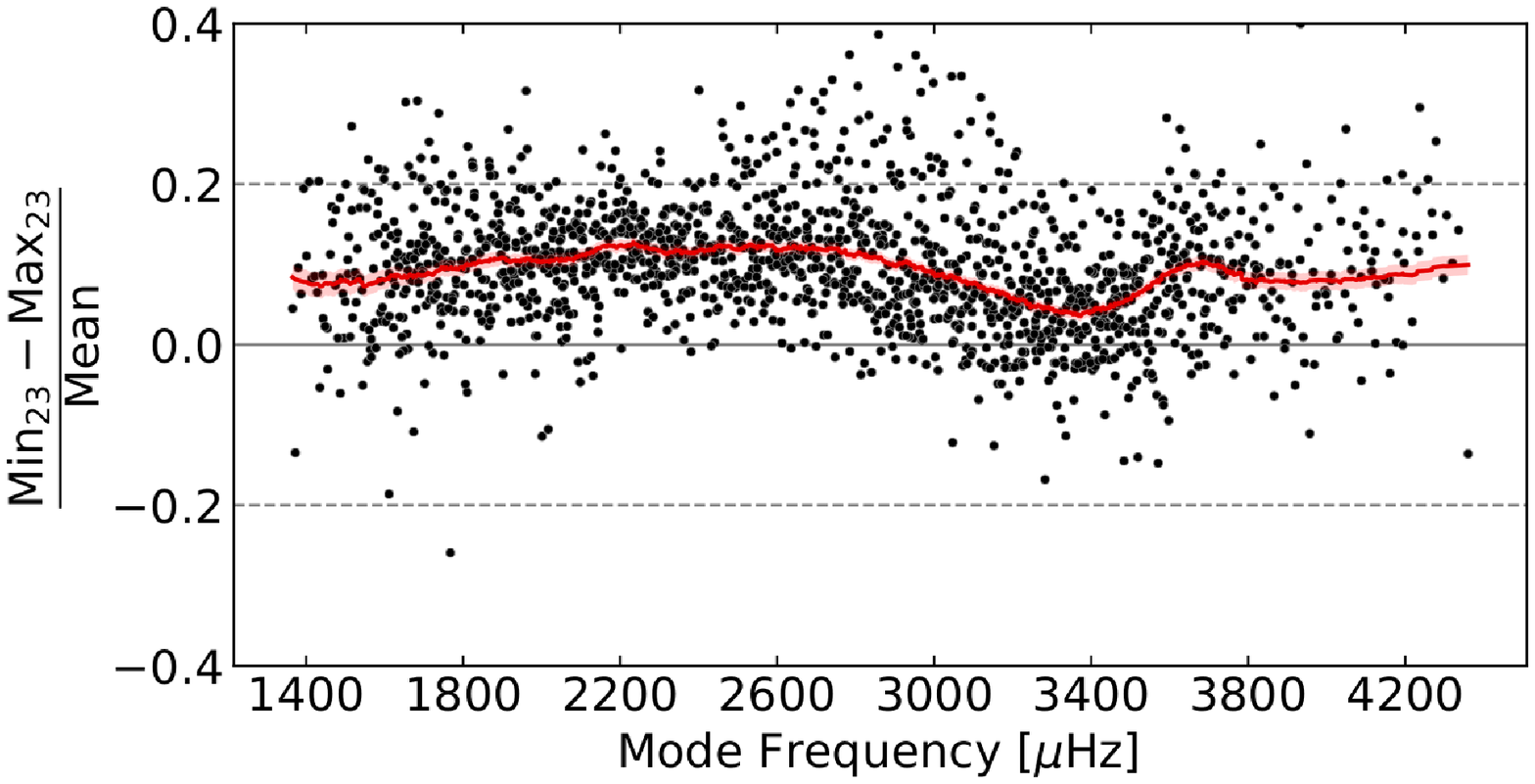}
	\includegraphics[width=0.49\linewidth]{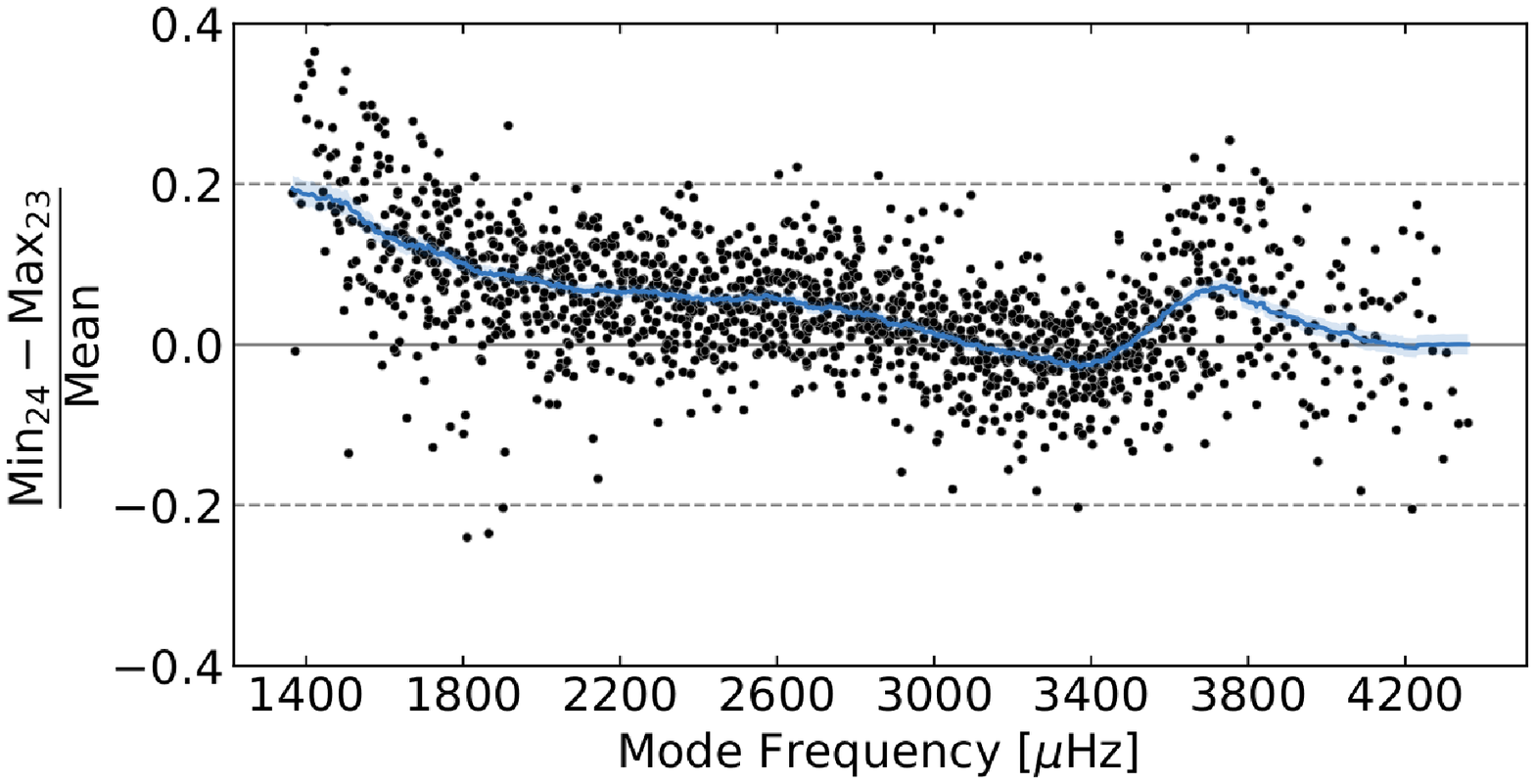}
	\includegraphics[width=0.49\linewidth]{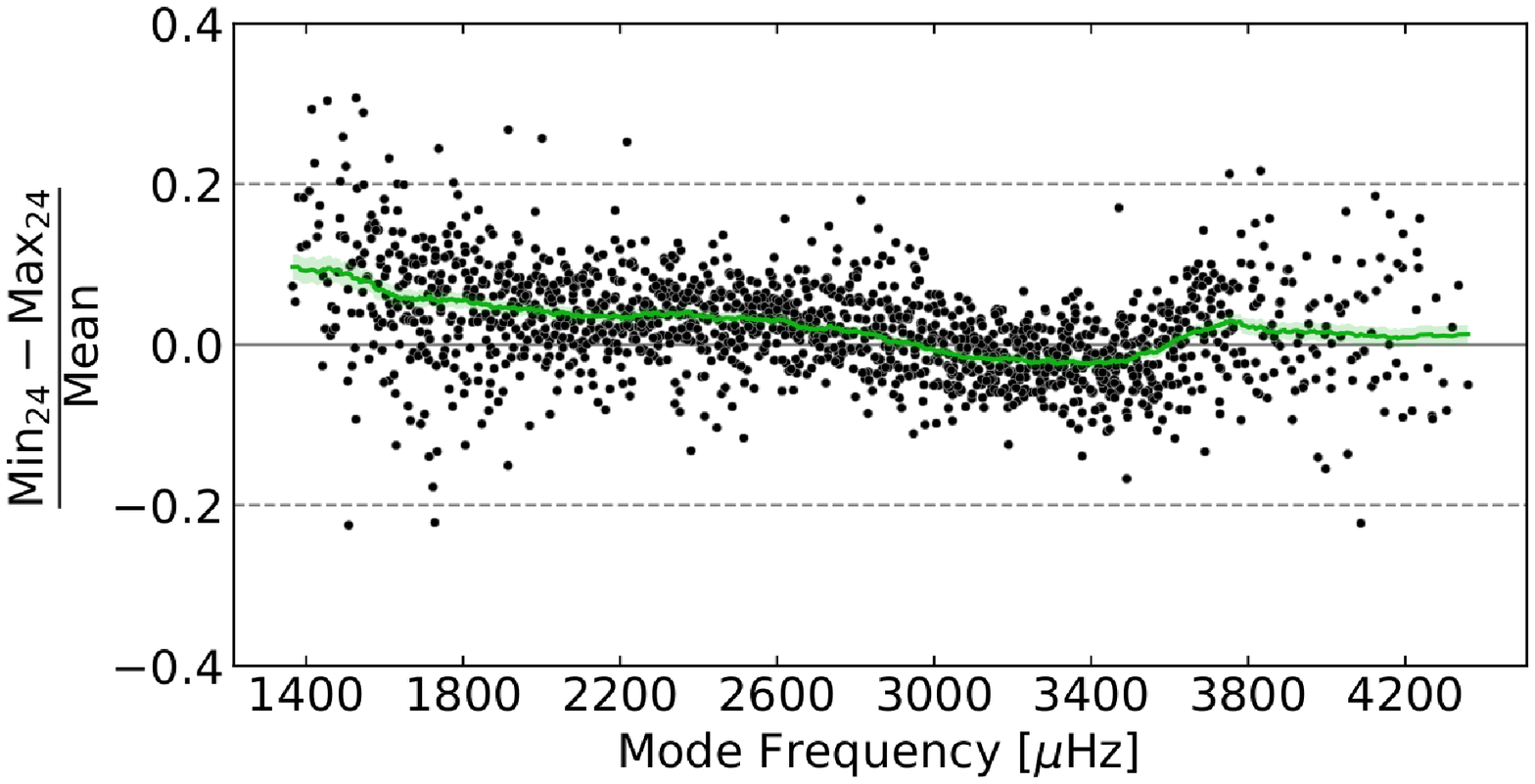}
	\includegraphics[width=0.49\linewidth]{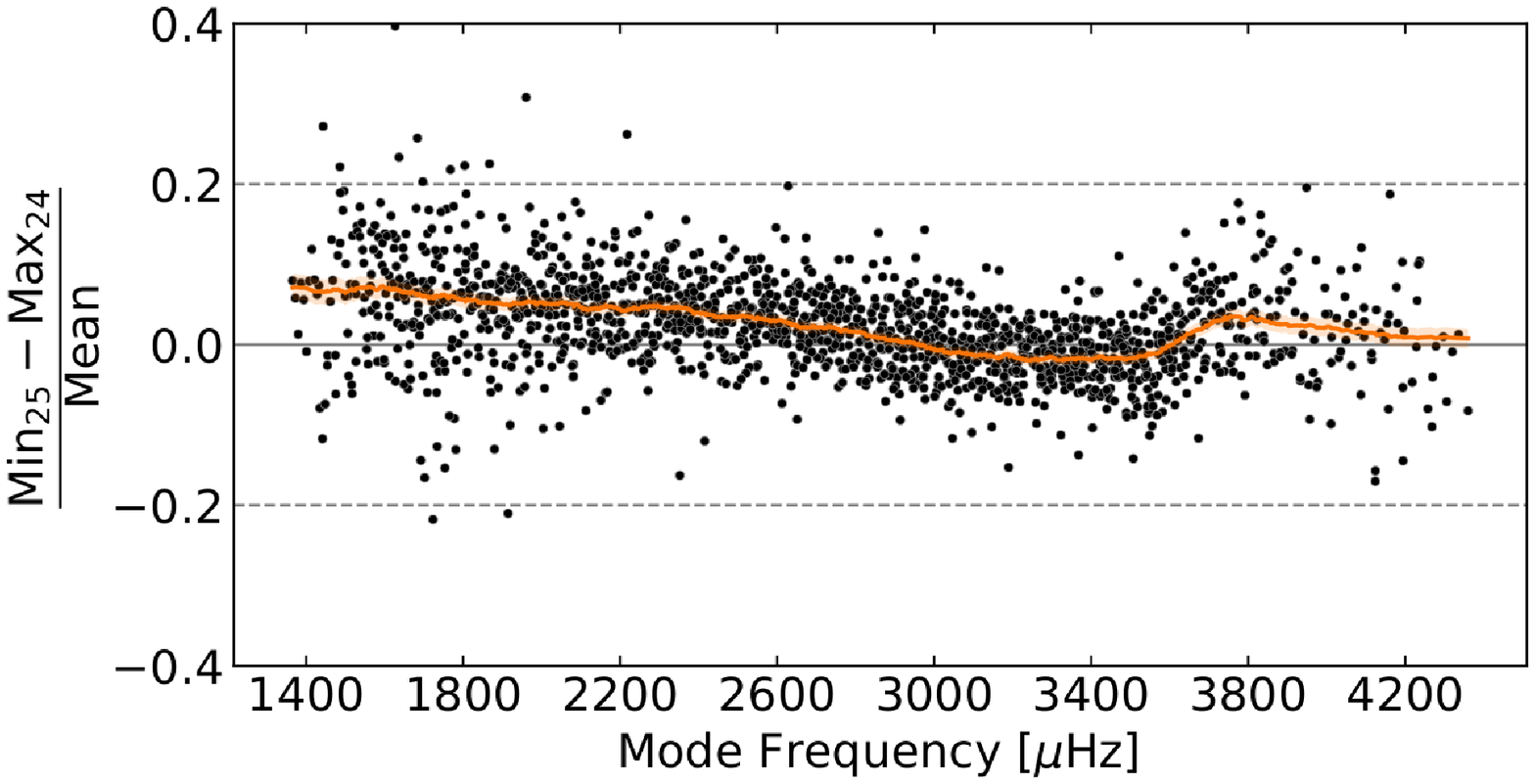}
	\includegraphics[width=0.49\linewidth]{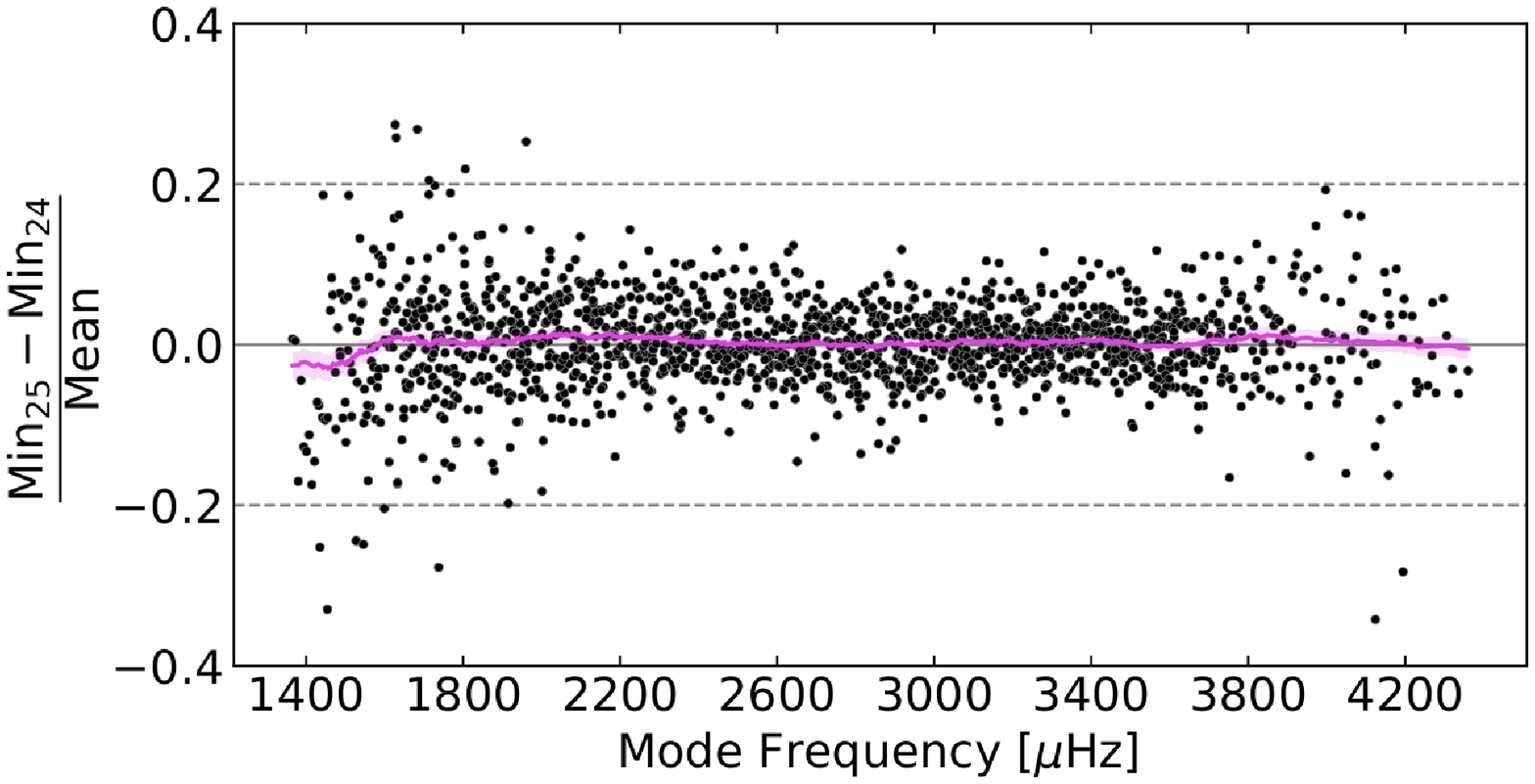}
	\includegraphics[width=0.49\linewidth]{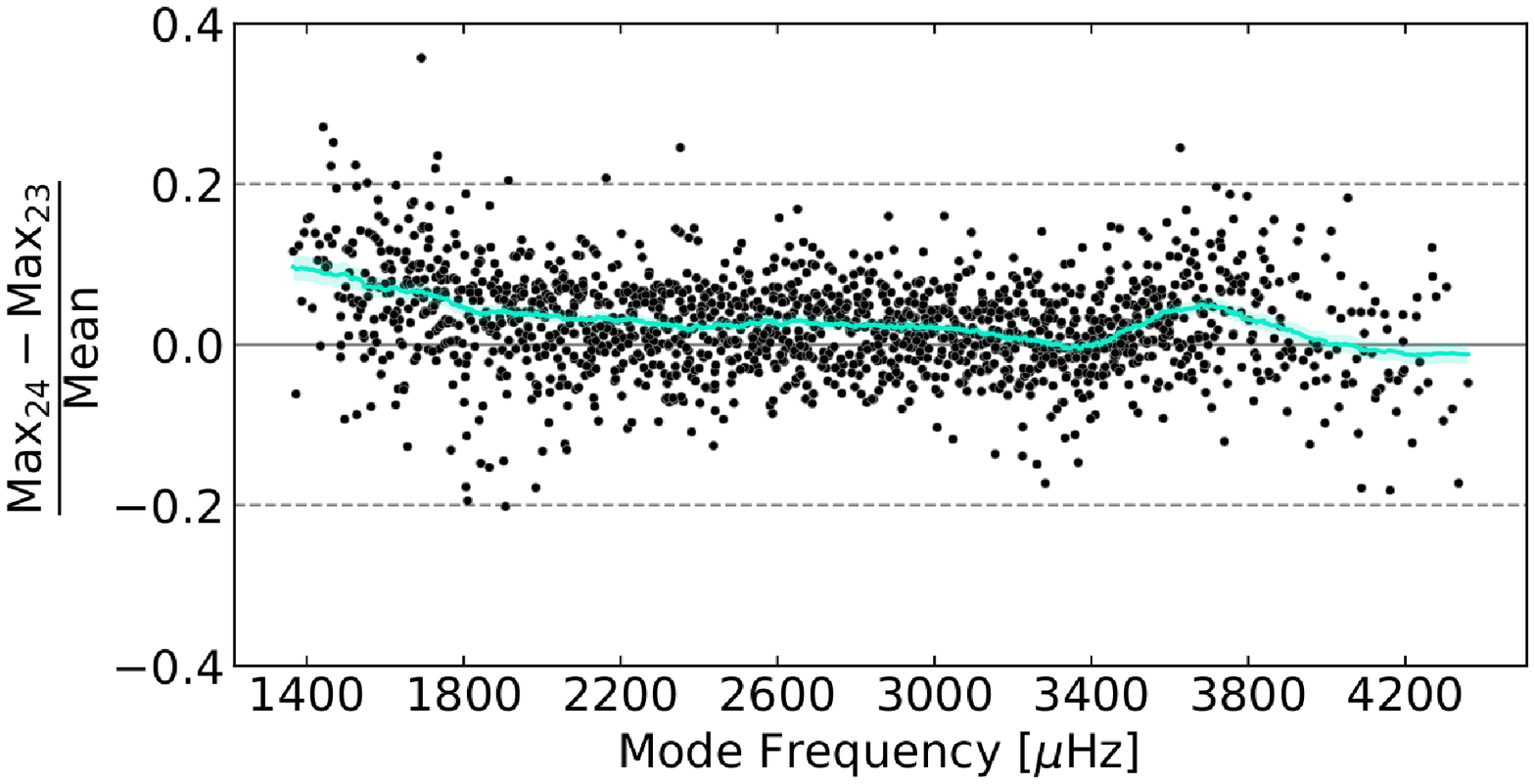}
	\caption{Difference between energy supply rates for individual modes for the activity extrema defined in Table~\ref{table:dates} as function of mode frequency. The variance-weighted average of each mode's energy supply rate is calculated over the extent of each extremal period. Differences are normalised by the mean over the entire GONG time series. The variance-weighted moving mean over 100 modes consecutive in frequency and its $10\sigma$ uncertainty are shown by the coloured lines and the shaded bands.}\label{fig:fracnu_unlagged}
\end{figure*}

\begin{table*}
	\caption{Differences in energy supply rates between the activity extrema stated in the first column for modes with frequencies stated in the top row. All values are given in \%. The dates of the activity extrema are taken from Table~\ref{table:dates}. Values are normalised by the mean over the complete time series, as in Figure~\ref{fig:fracnu_unlagged}. Then, the variance-weighted average over all modes in the parameter range is taken.}\label{table:extr_change_freq_org}
	\centering
	\begin{tabular}{ccccc}
		\hline
		&  \unit[1365--2400]\si{\micro\hertz} & \unit[2400--3300]\si{\micro\hertz}  &    \unit[3300--4360]\si{\micro\hertz}     &  \unit[1365--4360]\si{\micro\hertz}     \T\B\\ 
		\hline\T Min$_{23}$-Max$_{23}$ &10.70$\pm$0.04 & 9.67$\pm$0.03 & 6.39$\pm$0.04 & 9.19$\pm$0.02\\
			Min$_{24}$-Max$_{23}$          &8.32$\pm$0.04 & 2.61$\pm$0.03 & 0.90$\pm$0.04 & 3.87$\pm$0.02 \\
			Min$_{24}$-Max$_{24}$          &4.41$\pm$0.04 & 0.61$\pm$0.03 & -0.74$\pm$0.04 & 1.36$\pm$0.02\\
			Min$_{25}$-Max$_{24}$          &5.12$\pm$0.04 & 0.74$\pm$0.03 & -0.42$\pm$0.04 & 1.70$\pm$0.02\\
			Min$_{25}$-Min$_{24}$          &0.71$\pm$0.04 & 0.12$\pm$0.03 & 0.34$\pm$0.04 & 0.34$\pm$0.02 \\
		\B Max$_{24}$-Max$_{23}$       &3.89$\pm$0.04 & 2.02$\pm$0.03 & 1.62$\pm$0.04 & 2.47$\pm$0.02 \\
\hline
	\end{tabular}
\end{table*}
}

\afterpage{
\begin{figure*}
	\centering
	\includegraphics[width=0.49\linewidth]{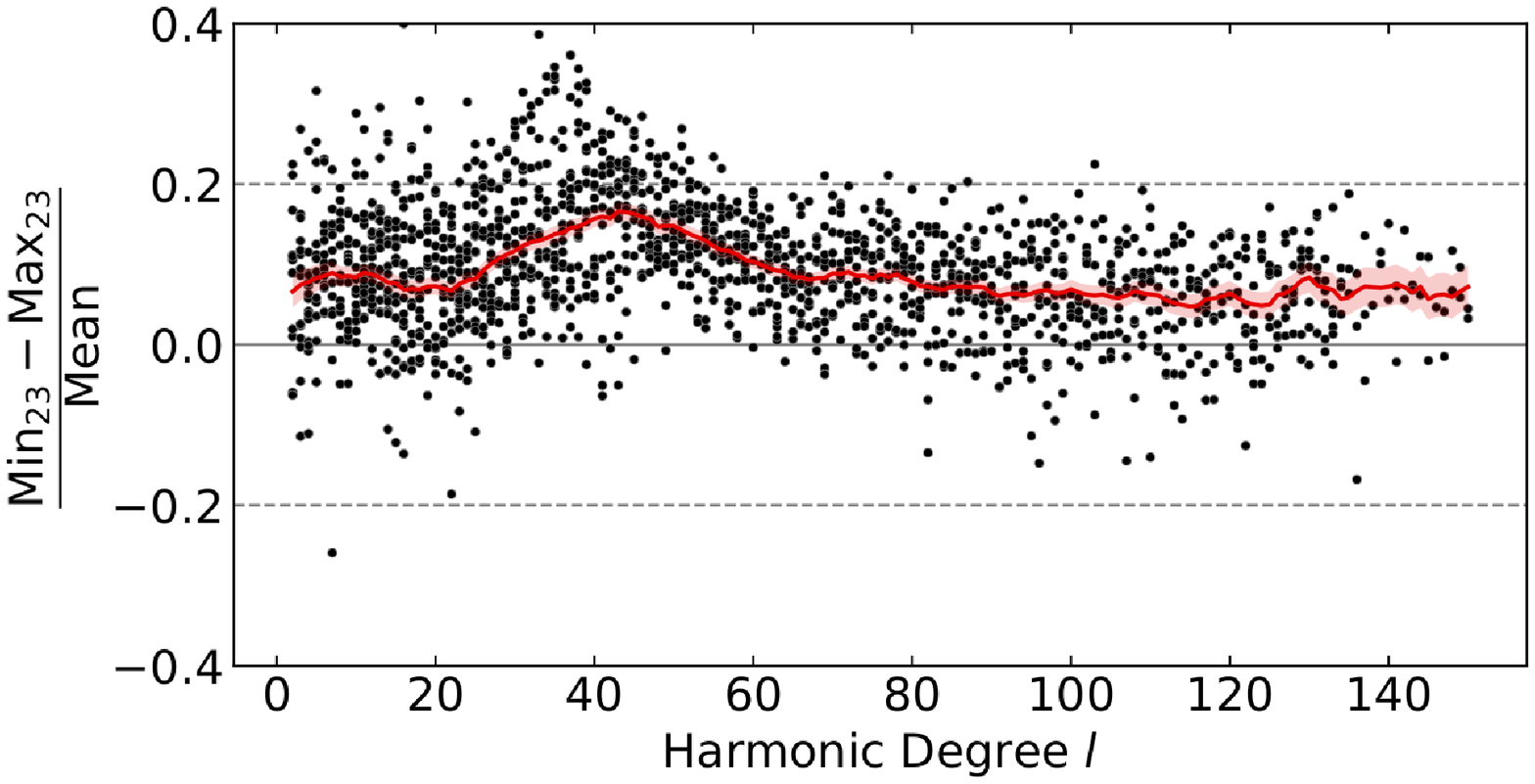}
	\includegraphics[width=0.49\linewidth]{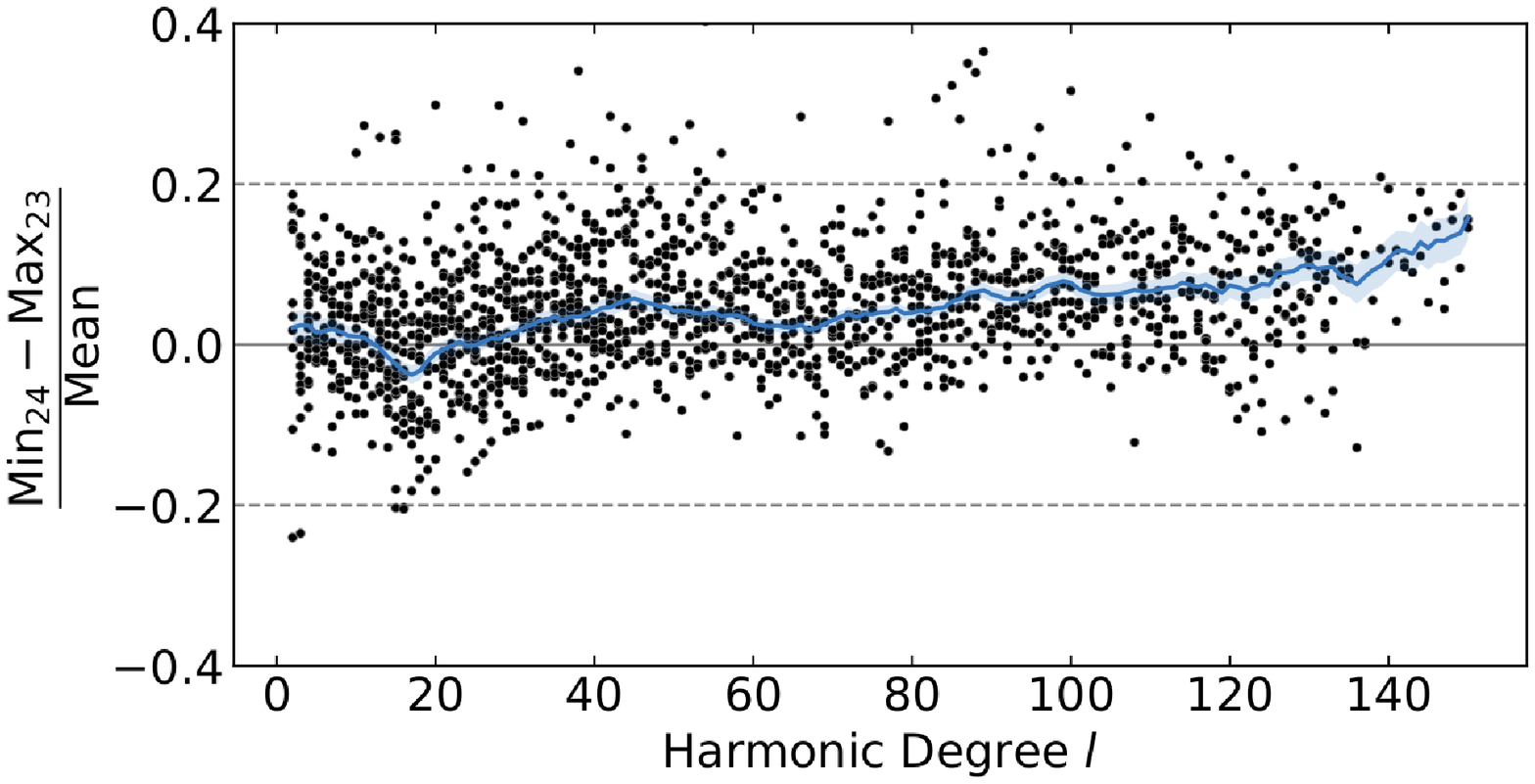}
	\includegraphics[width=0.49\linewidth]{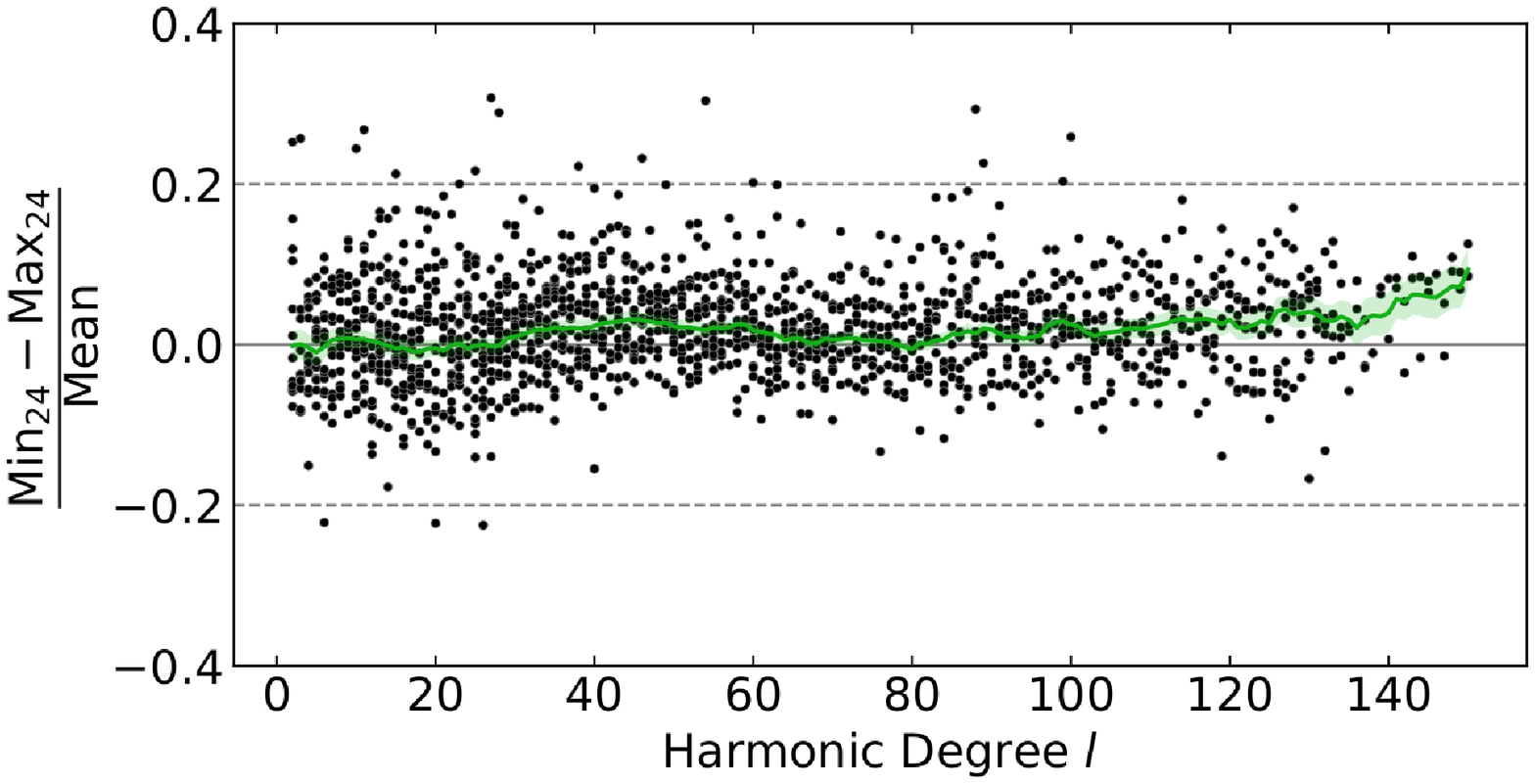}
	\includegraphics[width=0.49\linewidth]{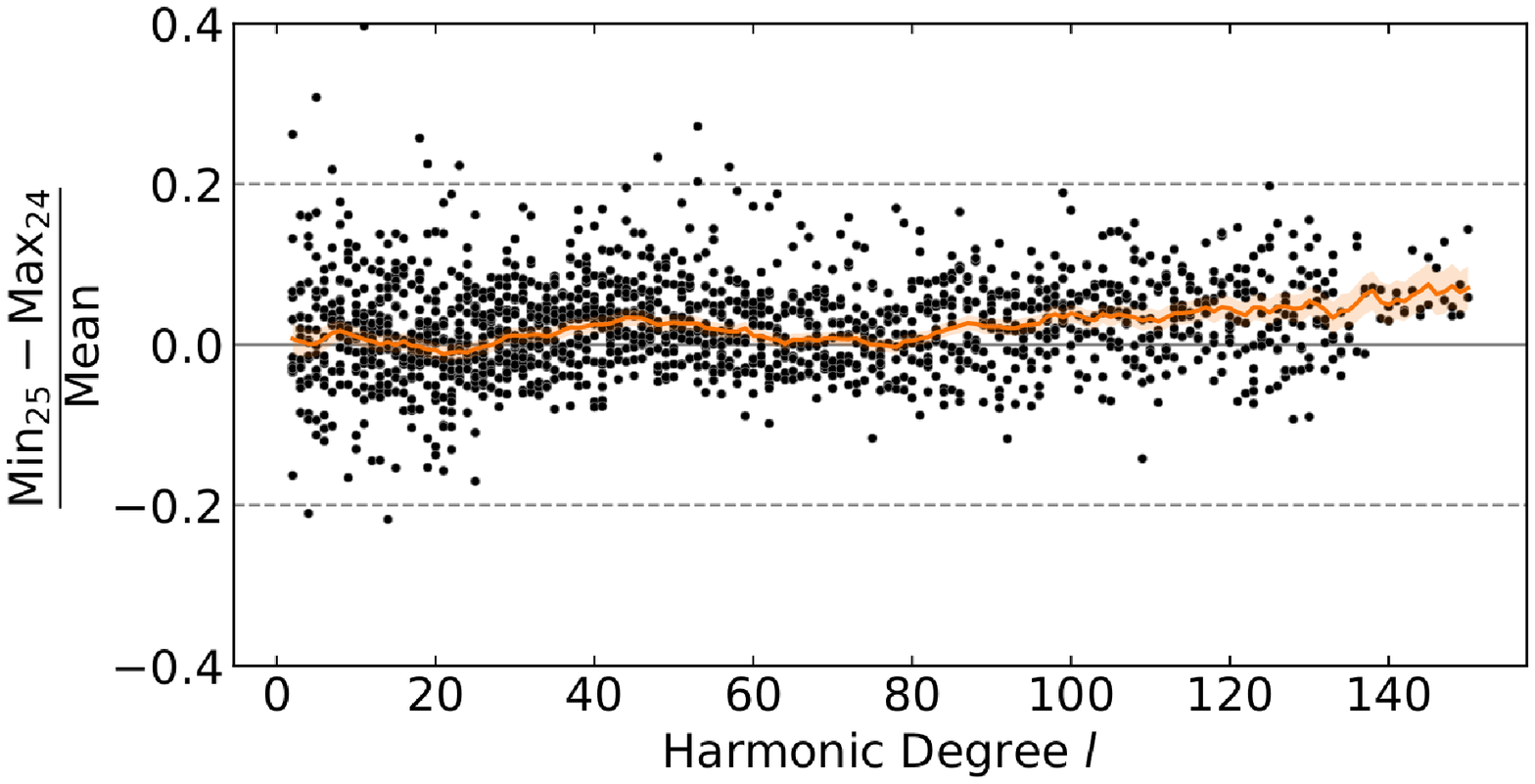}
	\includegraphics[width=0.49\linewidth]{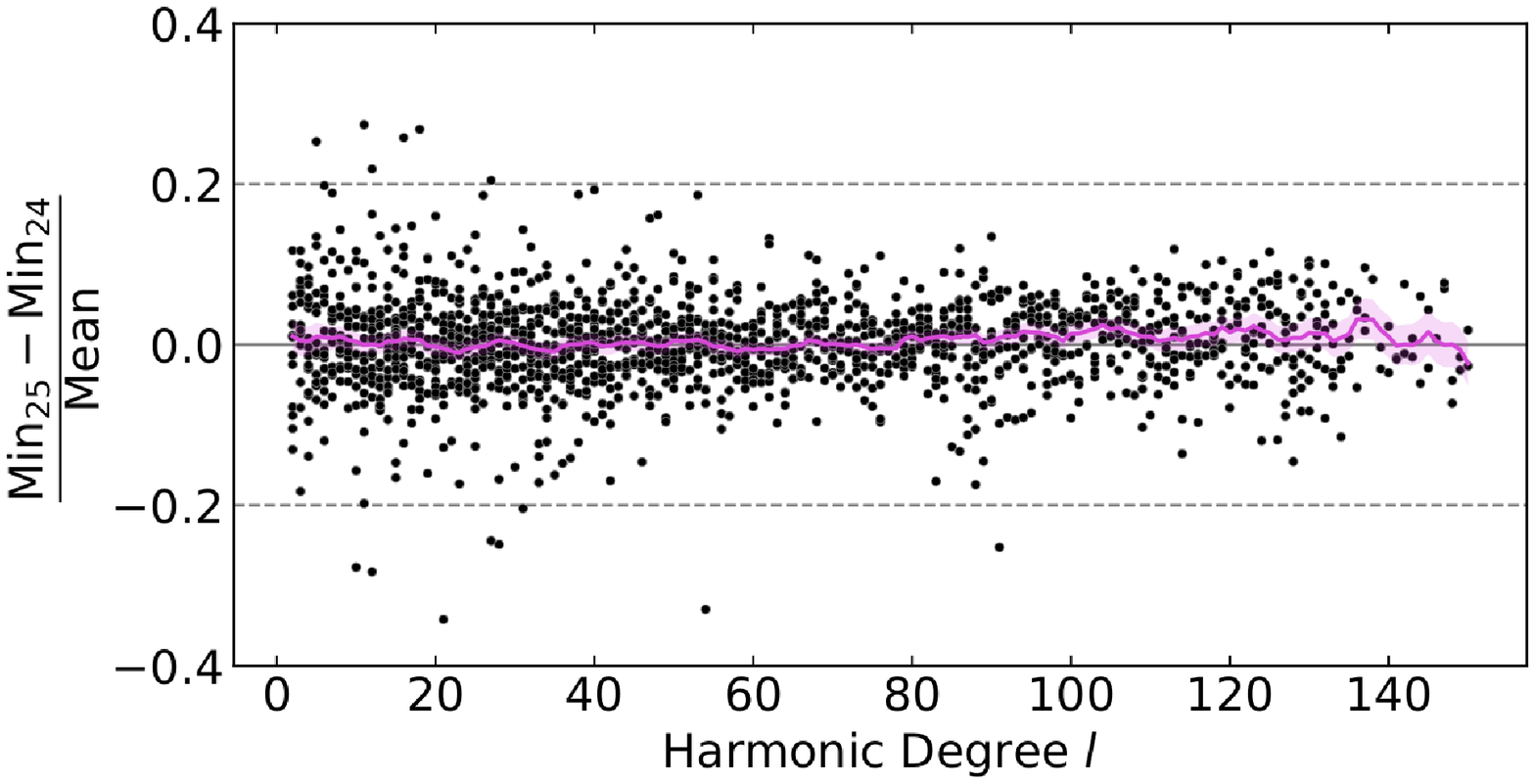}
	\includegraphics[width=0.49\linewidth]{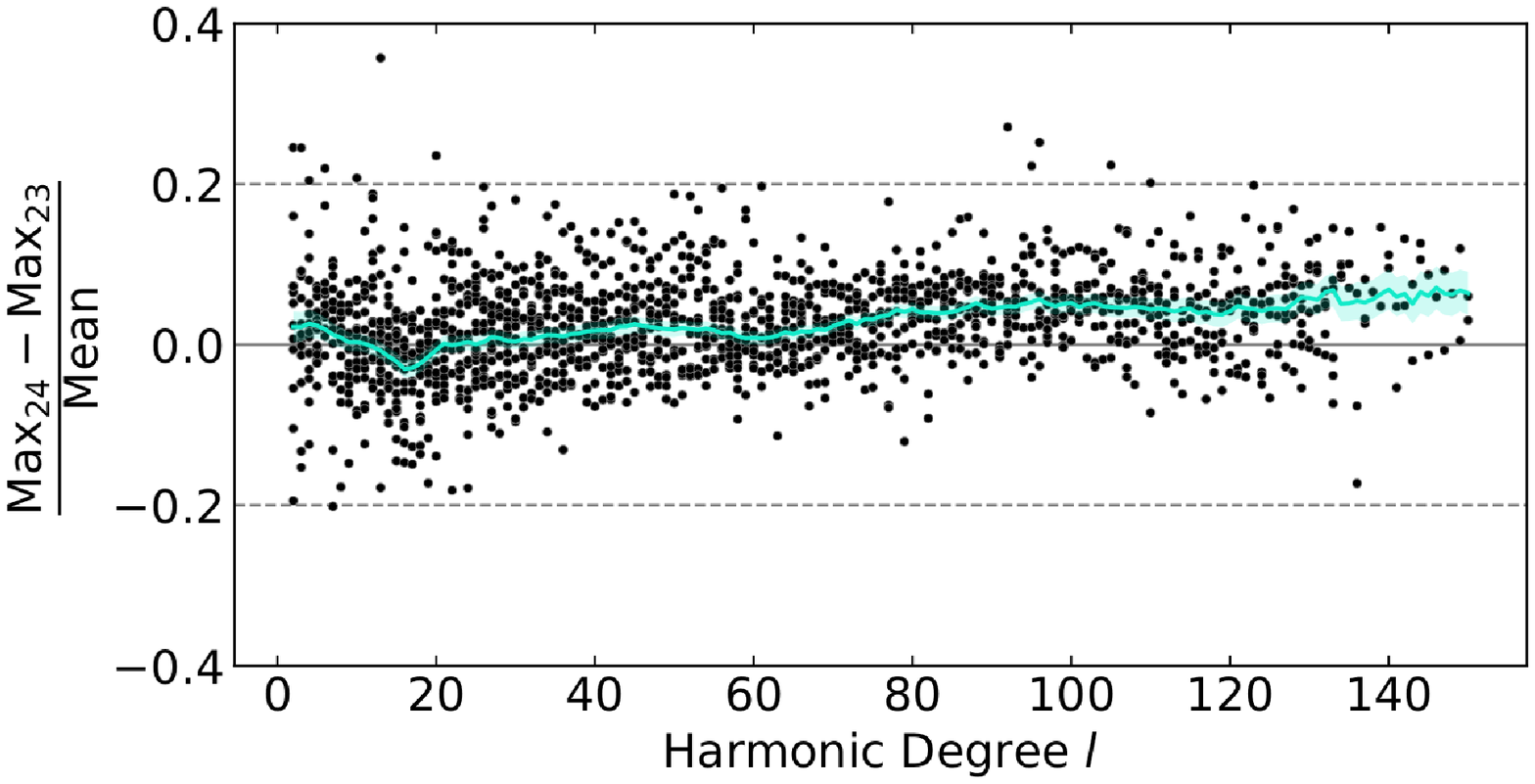}
	\caption{As Fig.~\ref{fig:fracnu_unlagged} but as a function of harmonic degree. The variance-weighted mean is calculated over all modes of five consecutive harmonic degrees. The weighted mean and its $10\sigma$ uncertainty are shown by the coloured lines and the shaded bands.}\label{fig:fracell_unlagged}
\end{figure*}

\begin{table*}
	\caption{Same as Table~\ref{table:extr_change_freq_org} but for mode sets from different harmonic degree ranges.}\label{table:extr_change_harmdeg_org}
	\centering
	\begin{tabular}{cccccc}
		\hline
		&  $l=2\text{--}30$ & $l=31\text{--}60$  &  $l=61\text{--}100$ & $l=101\text{--}150$                  &        $l=2\text{--}150$           \T\B\\ \hline
		\T 	Min$_{23}$-Max$_{23}$& 8.44$\pm$0.05 & 13.69$\pm$0.04 & 7.89$\pm$0.03 & 6.11$\pm$0.05 & 9.19$\pm$0.02 \\
Min$_{24}$-Max$_{23}$& -0.13$\pm$0.05 & 3.93$\pm$0.04 & 4.27$\pm$0.03 & 7.74$\pm$0.05 & 3.87$\pm$0.02\\
Min$_{24}$-Max$_{24}$& -0.04$\pm$0.05 & 2.33$\pm$0.04 & 0.88$\pm$0.03 & 2.79$\pm$0.05 & 1.36$\pm$0.02\\
Min$_{25}$-Max$_{24}$& 0.13$\pm$0.05 & 2.15$\pm$0.04 & 1.24$\pm$0.03 & 4.10$\pm$0.05 & 1.70$\pm$0.02 \\
Min$_{25}$-Min$_{24}$& 0.17$\pm$0.05 & -0.18$\pm$0.04 & 0.37$\pm$0.03 & 1.34$\pm$0.05 & 0.34$\pm$0.02\\
\B  Max$_{24}$-Max$_{23}$& -0.12$\pm$0.05 & 1.56$\pm$0.04 & 3.36$\pm$0.03 & 4.88$\pm$0.05 & 2.47$\pm$0.02 \\
\hline
	\end{tabular}
\end{table*}
}


\section{Discussion}\label{sec:6}
\subsection{Variation of supply rates}
\subsubsection*{Dependence on mode frequency and harmonic degree}
The energy supply rates of solar p modes vary over the solar activity cycle. The magnitude of this variation and the correlation with the $\overline{F_{10.7}}$ index varies with mode frequency and harmonic degree.

From Fig.~\ref{fig:matrix} and Table~\ref{table:amp} we find that energy supply rates of modes in the range $l=61\text{--}100$ with frequencies $<\SI{2400}{\micro\hertz}$ vary most strongly with a minimum of \SI{87.5(1)}{\percent} during times of high activity and a maximum of \SI{105.9(1)}{\percent} at low activity levels. The anti-correlation of this mode set with activity is strong with $r(0)=-0.75$ and $p<10^{-3}$. The strongest correlation is found for the mid-frequency mode set with $l=101\text{--}150$, for which $r(0)=-0.82$ and $p<10^{-3}$.

The amplitude of the variation over the activity cycle is smaller for higher frequency modes, which can be appreciated from Fig.~\ref{fig:matrix}, Table~\ref{table:amp}, and particularly from Fig.~\ref{fig:diff_fraction}. The anti-correlation at zero lag between these high frequency modes and $\overline{F_{10.7}}$ has $p<0.05$ only for the range $l=31\text{--}60$ and the full range of harmonic degrees. 

Interestingly, the strongest -- and most significant -- anti-correlation values for all mode sets except the highest harmonic degrees $l=101\text{--}150$ are found at positive lag values. I.e., it takes the energy supply rates a certain amount of time to react to changes in the level of magnetic activity. 

\subsubsection*{Why has this not been detected before?}
As the detected variation of $dE\slash dt$ over the activity cycle is rather large and very significant for the majority of mode sets, it is worthwhile to ask why this variation has not been detected until now. 

Previous studies of the solar-cycle variation of p-mode parameters often focused on mode frequencies (see Section~\ref{sec:1} for references), as these hold the most accessible information about the deep layers of the Sun. When studies did include the other mode parameters -- damping width $\Gamma$, mode amplitude $A$, and quantities compound out of these two -- they were almost exclusively limited to harmonic degrees $l\leq2$. The total number of azimuthal orders per radial order of the multiplets $l=0,1,2$ is 9. What is more, this effectively reduces to 6, as only modes for which $l+m$ is even are clearly seen in Sun-as-a-star data. Thus, even in the case if all harmonic azimuthal orders are fitted equally precisely \citep[which is not the case for unresolved observations e.g.,][]{2004A&A...424..713C}, this is equivalent to using only the $l=3$ modes from our study and averaging over their azimuthal orders. Mind that Figure~\ref{fig:lowl} in fact includes $l=2\text{--}4$ from resolved observations, i.e., 21 azimuthal orders per radial order.

\begin{figure*}
	\includegraphics[width=\linewidth]{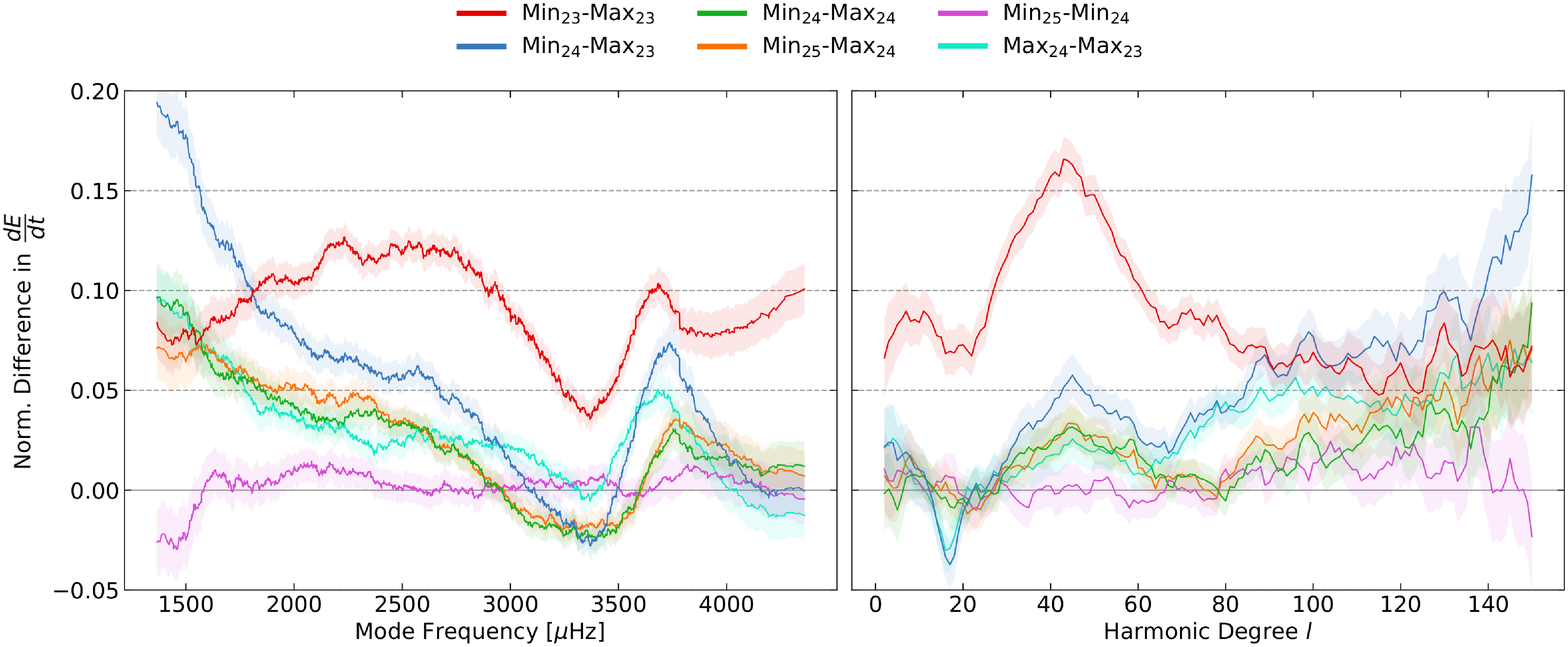}\caption{Smoothed differences between the energy supply rates of the extrema of magnetic activity normalised by the mean over the entire GONG time series as shown in Figs.~\ref{fig:fracnu_unlagged} and \ref{fig:fracell_unlagged}. Colours indicate the pairs of activity extrema that are subtracted from each other as given at the top.}\label{fig:diff_fraction}
\end{figure*}

In addition, it can be seen from the right panel of Figure~\ref{fig:diff_fraction} that lower harmonic degrees have smaller normalised differences between activity extrema. Thus, these modes' $dE\slash dt$ appear to be less sensitive to activity compared to modes of higher degree. The larger deviations from unity around the end of Max$_{\text{23}}$ and just before the beginning of Min$_{24}$ seen in Figure~\ref{fig:lowl} can also be found in the larger mode sets in Figure~\ref{fig:matrix}: low-degree modes do appear to have at least some response to the changing level of activity. Overall, the Pearson correlation-coefficient $r$ at zero lag between the normalised differences of the $l=2\text{--}4$ mode set plotted in Figure~\ref{fig:lowl} and the $\overline{F_{10.7}}$ index is only $r(0)=-0.18$ with a $p$-value of 0.1.

From this we conclude that some combination of a smaller number of multiplets in total, the small number of azimuthal orders to average over, the shorter length of the observations, and the fact that lower harmonic degrees appear to be less sensitive to magnetic activity, lead to the null results of earlier studies on the matter of activity-related variations of $dE\slash dt$.

\subsubsection*{Sensitivity of $dE\slash dt$ to magnetic activity}
\begin{figure}
	\includegraphics[width=\linewidth]{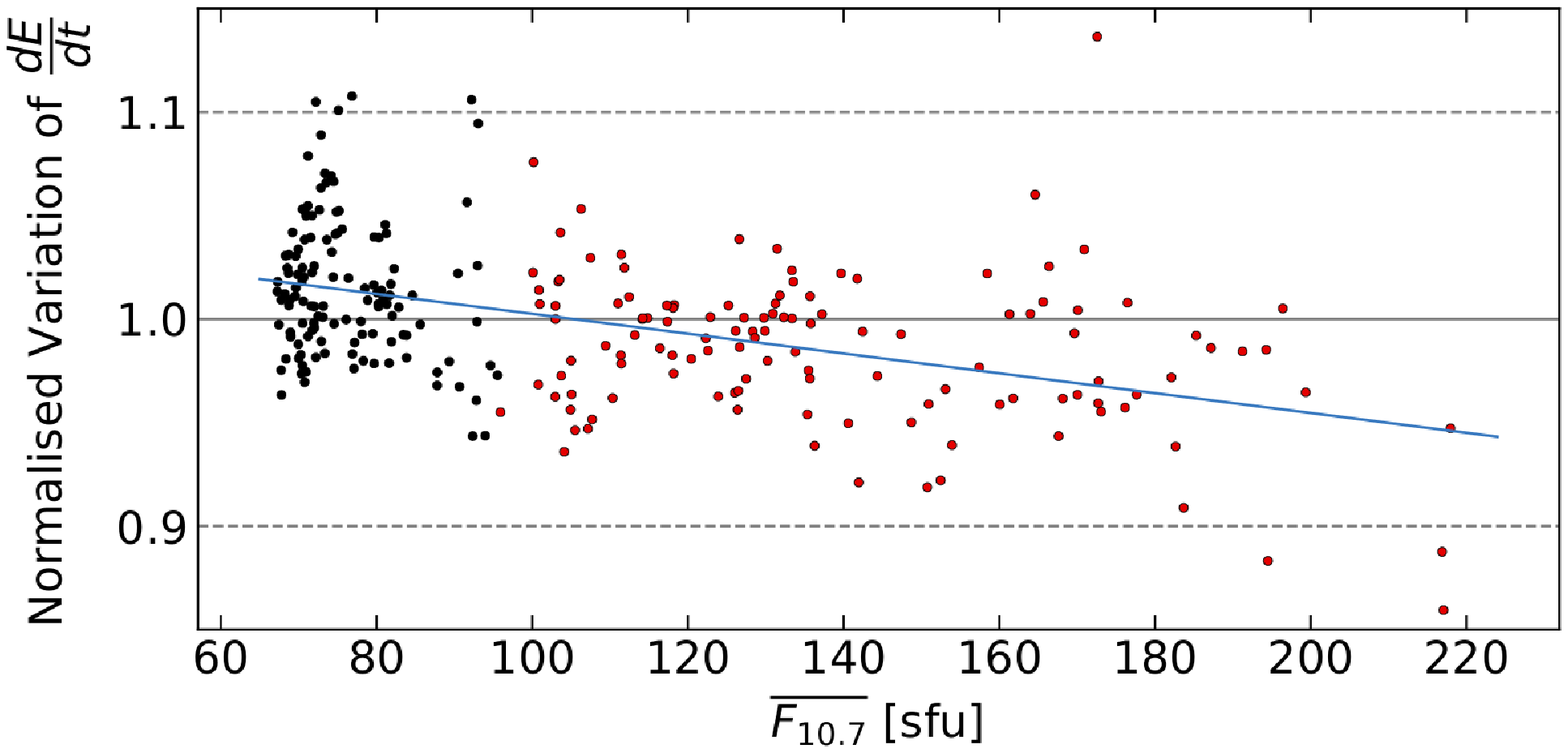}\caption{Normalised variation of the full mode set, which includes all frequencies and harmonic degrees, as a function of the $\overline{F_{10.7}}$ index. The solid blue line is a linear regression of the data. }\label{fig:dEdt_vs_F107} 
\end{figure}

Figure~\ref{fig:dEdt_vs_F107} shows the normalised variation of $dE\slash dt$ for the full mode set (average over all frequencies and harmonic degrees) as a function of $\overline{F_{10.7}}$. From a linear fit (blue line) to this we find that 
\begin{align}\label{eq:linear_fit}
\left[dE\slash dt\right] = \left(1.050\pm0.007\right) - \left(4.8\pm 0.6\right)\cdot10^{-4}  \si{sfu^{-1}}\overline{F_{10.7}},
\end{align}
where the square brackets indicate that the energy supply rates are normalised to the mean over the full cycle and the entire mode set is included. Again, the linear fit shows the anti-correlation of the energy supply rates with activity: The quiet Sun supply rates are $>1$ and the slope is $<0$. 

Many of the data points above the linear fit belong to the rising activity phase going from minimum 23 to maximum 23. Whether this is part of a hysteresis pattern in $dE\slash dt$ over a full magnetic cycle or an artefact from the GONG data and the GONG hardware upgrade just at the end of the maximum of cycle 23 should be tested with data from other instruments. The much larger percental change in $dE\slash dt$ between the minimum and maximum of cycle 23 compared to the later extrema can also be seen in Tables~\ref{table:extr_change_freq_org} and \ref{table:extr_change_harmdeg_org}. When comparing Min$_{24}$ to Min$_{23}$ directly, we find that the supply rates of the full mode set are \SI{5.51(2)}{\percent} lower for Min$_{24}$, which is obviously in contrast to the surface tracers of activity, e.g., the sunspot cycle.
It is also conceivable that, due to the different strength of cycle 24, the energy supply rates have settled at a slightly different level when compared to the earlier stronger cycles which had larger amplitudes between their activity minima and maxima.

As Figure~\ref{fig:fracnu_unlagged} and Table~\ref{table:extr_change_freq_org} show, modes around and above the Sun's frequency of maximum oscillation amplitude $\nu_{\text{max}}$ ($\approx\SI{3080}{\micro\hertz}$, e.g., \citealt{Kiefer2018a}) are less sensitive to magnetic activity than modes below $\nu_{\text{max}}$. Interestingly, this is in contrast to mode frequencies for which the amplitude of change along the cycle increases with mode frequency \citep[e.g.,][]{Broomhall2017}. Furthermore, let us consider the results of \cite{Kiefer2018a}, in particular their figures 6 and 7 which show the analogons of our Figure~\ref{fig:matrix} but for mode widths $\Gamma$ and mode amplitudes $A$. While \cite{Kiefer2018a} did not compare activity extrema separately, it can be appreciated that the percental variation in the low frequency mode set going from one activity extremum to the next is larger for $dE\slash dt$ than it is for $\Gamma$ or $A$. Also, the variation is larger for the low frequency mode set than the mid frequency set for $dE\slash dt$, but for $\Gamma$ and $A$ separately, the variation is larger in the mid frequency set. This is most apparent for the damping widths $\Gamma$ in figure 6 of \cite{Kiefer2018a}. All this shows that it is indeed not an increase in the damping alone that is causing the variations in $\Gamma$ and $A$ as has been assumed thus far. Indeed these observations call for an activity related variation in the forcing function of solar p modes.

\subsubsection*{Lag between $dE\slash dt$ and $\overline{F_{10.7}}$}
As we have established, the high frequency mode set is least reliable for the purpose of investigating $dE\slash dt$; so let us exclude this set from the considerations of this subsection and focus on the other three frequency ranges. 

Looking at the minima of the CCFs between $dE\slash dt$ and $\overline{F_{10.7}}$ and the minima of the fits to them, we find that the lag value for these decreases with increasing harmonic degree. For example, in the mid frequency range, the position of the global minimum of the CCF decreases from \SI{864}{\day} for modes in the range $l=2\text{--}30$, to \SI{756}{\day} for $l=31\text{--}60$, to \SI{108}{\day} for $l=61\text{--}100$, to finally \SI{0}{\day} for modes with $l=101\text{--}150$. This is interesting as the higher harmonic degree modes are confined to more shallow layers than modes of lower harmonic degrees \citep{Basu2016}. It thus appears that surface activity affects those modes earlier which are confined closer to the surface and modes that penetrate deeper into the Sun are fully affected later. As p modes are assumed to be excited very close to the surface \citep[e.g.,][]{Chaplin1999z} it is not immediately apparent why low degree, high inertia modes should take longer to react to changes in surface activity. We speculate that the higher mode inertia of lower frequency and lower harmonic degree modes acts as a sort of buffer against changes in the mode forcing, i.e., a stronger or longer-lasting perturbation is needed to induce a change in their energy supply rates.

We further note the curious behaviour in all mode sets' CCF just after the zero lag: compared to the polynomial fit, the CCF dips to smaller correlation values at around a lag of \SI{108}{\day}. It appears to oscillate back above the fit only to decrease again slightly. We speculate that this \mbox{(quasi-)oscillatory} behaviour in the CCF might be connected to the Sun's quasi-biennial variations that have been observed in, e.g., p-mode frequencies \citep{Broomhall2012}. This dip persists even if all time samples are included in the calculation of the cross-correlation functions. This feature is thus likely not caused by the time resolution of \SI{108}{\day} used in Fig.~\ref{fig:ccf}.

\subsubsection*{Quasi-yearly variation of $\Theta$}
Even after removing the residual change with apparent solar radius, the energy supply parameter $\Theta$ in Figure~\ref{fig:map} shows a quasi-yearly variation that needs to be further investigated in future studies. We attempted to remove this variation by fitting $\Theta_{nl}$ as a function of the solar position angle $P$ and the $B_0$ angle. However, whatever small variation with these quantities were removed by the linear fit, the observed variation in $\Theta$ persisted. We thus refrained from considering this matter further and call for a more detailed investigation of this (quasi-)yearly variation. We have not investigated periodograms of the variation of $dE\slash dt$. It is therefore entirely possible that the short term variation that can so clearly be seen in Figure~\ref{fig:map} has a period that is quite different from \SI{1}{yr} and might be connected to the above mentioned quasi-biennal oscillations. Part of the residual seasonal variation may be connected to changes in the GONG leakage matrix \citep{Hill1998}.

\subsection{Impact on asteroseismology}
The accurate prediction of the amplitudes of stellar p-mode oscillations is crucial in the target selection of photometric space missions. Such a target selection has been done for the asteroseismic targets of the NASA Transiting Exoplanet Survey Satellite mission \citep[TESS,][]{Ricker2014} based on the results of \cite{Campante2016} who calculated a detection probability for solar-like oscillations. Their equation for the prediction of the oscillation amplitudes in the power spectrum contains factors that account for the length of the data strings, sampling and instrumental effects, the pattern of p-mode peaks in the power spectrum and a factor that gives the maximum oscillation amplitude of radial modes. The calculation of this amplitude factor is based on scaling relations based on the work of \cite{Kjeldsen1995} and \cite{Chaplin2011}. \cite{Campante2016} mention the need for accounting for the effect of magnetic activity in the estimation of mode amplitudes. 

During the \textit{Kepler} mission \citep{Koch2010, Borucki2010}, over 2000 stars were observed in the survey phase. Of these, only 540 were detected to exhibit solar-like oscillations \citep{Mathur2019}. It is known that -- as in the solar case -- stellar oscillations are suppressed by magnetic activity \citep[e.g.,][]{Garcia2010, Salabert2017a, Kiefer2017, Santos2018}. Indeed, \cite{Mathur2019} were able to attribute \SI{32}{\percent} of non-detections to magnetic activity. The remaining \SI{68}{\percent} however are still to be accounted for. 

Based on the results we present in this article, we conjecture that it is not only increased damping of oscillations and the ensuing attrition of mode amplitudes that impede the detection of stellar p-modes and need to be accounted for, but in fact the oscillations are fed with less energy per unit time even for moderately active stars like the Sun.

In the target selection of future asteroseismic surveys this should be taken into account for stars where a reliable measure of activity is available. If the detection probability of oscillations is to be maximised, more active stars ought to be deprioritised. More work on this has to be done, e.g., on the extent to which the lowest harmonic degrees are affected by activity. This problem could be tackled by careful analyses of the available photometric time series on the one hand \citep{Salabert2017a} and simulations of mode excitation in solar and stellar convection zones \citep[e.g., ][]{Belkacem2006, Samadi2007, Zhou2019, 2020MNRAS.495.4904Z} that investigate the dependence of the mode excitation rate in these simulations on the magnetic field strength on the other hand.

\subsection{Shortcomings}\label{sec:6.3}
We have corrected the jumps in mode amplitudes that were noticed by \cite{Kiefer2018a} in the same empirical way they used: we applied a correction factor to remove the two obvious jumps (see Section~\ref{sec:31}). This is unsatisfactory, as neither the origin of the second jump is explained, nor can we be certain that the direction of the adjustment is correct. It is entirely possible that the absolute values of the mode energy supply rates as presented in Figures~\ref{fig:erate_ellnu}, \ref{fig:erate_avg} and Table~\ref{table:erates} are underestimated, as the correction factor assumes the time between 2001 and 2005 as the baseline. Data before the jump in 2001 are corrected downwards, data after the 2005 jump are corrected upwards. All our time dependent analyses are unaffected by this choice of a baseline, only statements about the absolute values of \mbox{$dE\slash dt$} are concerned with this.

We have neglected possible contributions from time-dependent mode asymmetry in this study. We are aware that some fraction of the detected activity-related change of supply rates can potentially be attributed to changes in mode asymmetry. With the publicly available official GONG data such analyses cannot be done, as the profiles that are used in the GONG mode fitting pipeline are symmetrical. It is therefore expedient to repeat similar analyses as we have done here with data that include a mode asymmetry parameter \citep[e.g.,][]{Korzennik2017} and, in particular, with data from the SOHO-MDI \citep{Domingo1995, Scherrer1995} and SDO-MDI \citep{Pesnell2012, Schou2012} instruments.

The GONG leakage matrix is not included in the fitting of the mode peaks. We account for part of this shortcoming by removing the bowl-shape of the energy supply rates as a function of $|m/l|$, see Section~\ref{sec:311}. This effectively takes care of the $m$-dependence of the self-leakage. However, the overall level of self-leakage is only naively corrected for by the inclusion on the factor $C_{\text{vis}}$ in Eq.~\ref{eq:dEdt}. We stress that using this factor for all harmonic degrees is consistent with \cite{Komm2000a}. In reality however, it can vary for different harmonic degrees \citep[see, e.g.,][]{Baudin2005}. We do not anticipate $C_{\text{vis}}$ to vary over the course of the Sun's activity cycle, thus it cannot contribute to the detected cycle variation – it only affects the absolute values of energy supply rates. In a numerical estimation of the geometrical visibilities of different harmonic degrees in resolved Doppler observations, we found that the self-leakage levels off at relatively low harmonic degree $l\approx 4$. Thus, the reported absolute averages of energy supply rates are most likely only off by a small constant factor, as the majority of modes have $l>4$.


\section{Summary and Conclusions}\label{sec:7}
We have, for the first time, analysed the solar p-mode energy supply rates up to $l=150$ through two solar activity cycles. Contrary to the long-held opinion that the energy that is fed into the p-modes per unit time is constant, we have found that they are indeed sensitive to the level of magnetic activity.

Averaged over the full mode set, the minimum supply rates are found during activity maxima, with a minimum of \SI{91.2(1)}{\percent} normalised to the entire time series. At activity minimum they reach as high as \SI{107.6(1)}{\percent}, i.e., supply rates are in anti-phase with the activity cycle with a correlation of $r=-0.50$ $(p<10^{-3})$, where the $F_{10.7}$ index was used as a proxy for solar activity.

The fractional change through the activity cycle depends somewhat on mode frequency and only weakly on harmonic degree: supply rates of lower frequency modes $<\SI{3000}{\micro\hertz}$ are more strongly affected than higher frequency modes. It appears that the supply rates of lower harmonic degree modes $l\lesssim20$ tend to be least affected by activity.

According to the mode energy supply rates, the currently ongoing minimum is as deep as the long minimum between cycles 23 and 24. Averaged over the full mode set, there is very little variation between these two minima with a difference of only \SI{0.33(2)}{\percent}. Also, it is reflected in the energy supply rates that the maximum of cycle 23 was stronger than that of cycle 24: the energy supply rates were \SI{2.43(2)}{\percent} higher during the latter.

As the energy supply rates are a crucial ingredient in simulations and prediction of the solar and stellar p-mode amplitudes, it is worthwhile to establish their ground-state, i.e., quiet Sun levels. We provide a table with the energy supply rates as a function of mode frequency for the average over the entire time series, corrected for mode inertia, averaged over the periods of activity minima as well as maxima, and with the mode inertiae normalised at both the photospheric radius $R_{\odot}$ as well as the GONG observation height $r_{\text{obs}}$. Table~\ref{table:erates} as well as a larger table with the energy supply rates of 1606 modes are provided online.

Our findings could -- and should -- be corroborated by analysing the last two activity cycles with data from multiple instruments. Optimally, such a study is done with results obtained with both symmetrical and asymmetrical mode profiles. This way, it could be determined to what extent the here reported change in supply rates over the cycle is in fact due to variations in mode asymmetry. Finally, we note that the convective forcing of acoustic mode in stars with different metallicity and surface gravity than the Sun might respond more weakly or strongly to magnetic activity. 

\section*{Acknowledgements}
R. K. and A-M. B. acknowledge the support of the Science and Technology Facilities Council (STFC) consolidated grant ST/P000320/1. A-M. B. acknowledges the support STFC consolidated grant ST/T000252/1. 

We thank Rudi Komm for supplying us with his code underlying \cite{Komm2000a}, which we used as a starting point for this work. The authors thank the anonymous referee for taking the time to review this article and for their useful comments.

This work utilises GONG data obtained by the NSO Integrated Synoptic Program (NISP), managed by the National Solar Observatory, the Association of Universities for Research in Astronomy (AURA), Inc. under a cooperative agreement with the National Science Foundation. The data were acquired by instruments operated by the Big Bear Solar Observatory, High Altitude Observatory, Learmonth Solar Observatory, Udaipur Solar Observatory, Instituto de Astrofísica de Canarias, and Cerro Tololo Interamerican Observatory. 

\textsc{software}: This work made use of the following software libraries not all of which is cited in the text:
\textsc{matplotlib} \citep{2007CSE.....9...90H}, \textsc{NumPy} \citep{Oliphant2006}, \textsc{pandas} \citep{Pandas2019}, \textsc{SciPy} \citep{2020SciPy-NMeth}, \textsc{seaborn} \citep{Seaborn2018}, \textsc{statsmodels} \citep{seabold2010statsmodels}, \textsc{SunPy} v1.1.2 \citep{stuart_j_mumford_2020_3735273}, an open-source and free community-developed solar data analysis Python package \citep{sunpy_community2020}, \textsc{uncertainties} \citep{uncertainties}.

\section*{Data Availability}
The data underlying this article are available from the GONG FTP server, at \url{ftp://gong2.nso.edu/TSERIES/v1f} and from the Natural Resources Canada FTP server \urlfour.

\bibliographystyle{mnras}

\begin{appendix}
\onecolumn
\section{Table of energy supply rates}
\begin{table*}
	\caption{Energy supply rates averaged over 32 modes consecutive in mode frequency; the last row only contains 12 modes. The first column gives the mean mode frequency of each bin. The second and third columns are energy supply rates without and with rescaling by $Q$ averaged over the entire time series with variance-weighting over the 32 modes. The values in the following columns are averaged over only the indicated periods of activity extrema, i.e., all activity minima or maxima as defined in Table~\ref{table:dates}, and are scaled by $Q$ as indicated. To maintain readability, uncertainties are rounded up to 0.001. More digits are given in the online version of this table.}\label{table:erates}
	\begin{tabular}{ccccccc}
		\hline
		Mode Frequency& $\dfrac{dE}{dt}$ & $\dfrac{dE}{dt}\cdot Q$ &  $\dfrac{dE}{dt}\left(\text{Min}\right)$ &   $\dfrac{dE}{dt}\left(\text{Max}\right)$ &  $\dfrac{dE}{dt}\cdot Q\left(\text{Min}\right)$ &   $\dfrac{dE}{dt}\cdot Q\left(\text{Max}\right)$\T \rule{0pt}{4ex}    \\
		$\left[\si{\micro\hertz}\right]$ & $\left[\SI{e22}{erg.s^{-1}}\right]$&$\left[\SI{e22}{erg.s^{-1}}\right]$&$\left[\SI{e22}{erg.s^{-1}}\right]$&$\left[\SI{e22}{erg.s^{-1}}\right]$ & $\left[\SI{e22}{erg.s^{-1}}\right]$&$\left[\SI{e22}{erg.s^{-1}}\right]$\T\B \\
		\hline
        \T      1449.29 & 0.105$\pm$0.001 & 0.037$\pm$0.001 & 0.109$\pm$0.001 & 0.106$\pm$0.001 & 0.038$\pm$0.001 & 0.037$\pm$0.001\\
        1548.62 & 0.139$\pm$0.001 & 0.055$\pm$0.001 & 0.144$\pm$0.001 & 0.140$\pm$0.001 & 0.056$\pm$0.001 & 0.054$\pm$0.001\\
        1616.52 & 0.177$\pm$0.001 & 0.074$\pm$0.001 & 0.183$\pm$0.001 & 0.177$\pm$0.001 & 0.076$\pm$0.001 & 0.073$\pm$0.001\\
        1673.87 & 0.215$\pm$0.001 & 0.094$\pm$0.001 & 0.223$\pm$0.001 & 0.214$\pm$0.001 & 0.097$\pm$0.001 & 0.094$\pm$0.001\\
        1724.75 & 0.276$\pm$0.001 & 0.120$\pm$0.001 & 0.285$\pm$0.001 & 0.274$\pm$0.001 & 0.123$\pm$0.001 & 0.119$\pm$0.001\\
        1775.14 & 0.289$\pm$0.001 & 0.149$\pm$0.001 & 0.299$\pm$0.001 & 0.289$\pm$0.001 & 0.154$\pm$0.001 & 0.149$\pm$0.001\\
        1833.03 & 0.354$\pm$0.001 & 0.189$\pm$0.001 & 0.367$\pm$0.001 & 0.350$\pm$0.001 & 0.195$\pm$0.001 & 0.187$\pm$0.001\\
        1889.43 & 0.430$\pm$0.001 & 0.242$\pm$0.001 & 0.446$\pm$0.001 & 0.425$\pm$0.001 & 0.250$\pm$0.001 & 0.240$\pm$0.001\\
        1942.92 & 0.553$\pm$0.001 & 0.303$\pm$0.001 & 0.573$\pm$0.001 & 0.547$\pm$0.001 & 0.313$\pm$0.001 & 0.300$\pm$0.001\\
        1994.13 & 0.657$\pm$0.001 & 0.376$\pm$0.001 & 0.677$\pm$0.001 & 0.649$\pm$0.001 & 0.388$\pm$0.001 & 0.372$\pm$0.001\\
        2041.96 & 0.761$\pm$0.001 & 0.454$\pm$0.001 & 0.784$\pm$0.001 & 0.749$\pm$0.001 & 0.468$\pm$0.001 & 0.447$\pm$0.001\\
        2089.88 & 0.894$\pm$0.001 & 0.542$\pm$0.001 & 0.923$\pm$0.001 & 0.878$\pm$0.001 & 0.559$\pm$0.001 & 0.533$\pm$0.001\\
        2136.39 & 1.058$\pm$0.001 & 0.658$\pm$0.001 & 1.094$\pm$0.001 & 1.041$\pm$0.001 & 0.680$\pm$0.001 & 0.648$\pm$0.001\\
        2188.75 & 1.214$\pm$0.001 & 0.799$\pm$0.001 & 1.254$\pm$0.001 & 1.190$\pm$0.001 & 0.824$\pm$0.001 & 0.783$\pm$0.001\\
        2241.29 & 1.476$\pm$0.001 & 0.987$\pm$0.001 & 1.533$\pm$0.001 & 1.448$\pm$0.001 & 1.023$\pm$0.001 & 0.969$\pm$0.001\\
        2290.98 & 1.742$\pm$0.001 & 1.177$\pm$0.001 & 1.807$\pm$0.001 & 1.717$\pm$0.001 & 1.219$\pm$0.001 & 1.160$\pm$0.001\\
        2338.77 & 2.033$\pm$0.001 & 1.410$\pm$0.001 & 2.103$\pm$0.001 & 2.002$\pm$0.001 & 1.457$\pm$0.001 & 1.393$\pm$0.001\\
        2384.46 & 2.357$\pm$0.001 & 1.657$\pm$0.001 & 2.447$\pm$0.001 & 2.323$\pm$0.002 & 1.716$\pm$0.001 & 1.637$\pm$0.001\\
        2436.51 & 2.670$\pm$0.001 & 1.901$\pm$0.001 & 2.744$\pm$0.001 & 2.624$\pm$0.002 & 1.954$\pm$0.001 & 1.872$\pm$0.001\\
        2488.52 & 3.085$\pm$0.001 & 2.221$\pm$0.001 & 3.185$\pm$0.002 & 3.016$\pm$0.002 & 2.286$\pm$0.001 & 2.178$\pm$0.001\\
        2538.01 & 3.594$\pm$0.001 & 2.571$\pm$0.001 & 3.715$\pm$0.002 & 3.509$\pm$0.002 & 2.652$\pm$0.001 & 2.521$\pm$0.002\\
        2586.67 & 4.060$\pm$0.001 & 2.867$\pm$0.001 & 4.190$\pm$0.002 & 3.977$\pm$0.003 & 2.956$\pm$0.002 & 2.820$\pm$0.002\\
        2632.23 & 4.368$\pm$0.001 & 3.214$\pm$0.001 & 4.512$\pm$0.002 & 4.266$\pm$0.003 & 3.308$\pm$0.002 & 3.151$\pm$0.002\\
        2676.88 & 4.814$\pm$0.001 & 3.580$\pm$0.001 & 4.950$\pm$0.003 & 4.718$\pm$0.004 & 3.681$\pm$0.002 & 3.524$\pm$0.003\\
        2725.07 & 5.655$\pm$0.001 & 4.052$\pm$0.001 & 5.819$\pm$0.003 & 5.540$\pm$0.004 & 4.166$\pm$0.002 & 3.979$\pm$0.003\\
        2775.65 & 5.817$\pm$0.002 & 4.458$\pm$0.001 & 6.000$\pm$0.003 & 5.701$\pm$0.004 & 4.585$\pm$0.002 & 4.382$\pm$0.003\\
        2824.68 & 6.868$\pm$0.002 & 4.955$\pm$0.001 & 7.066$\pm$0.004 & 6.704$\pm$0.005 & 5.082$\pm$0.003 & 4.856$\pm$0.003\\
        2872.46 & 7.090$\pm$0.002 & 5.467$\pm$0.002 & 7.257$\pm$0.004 & 6.945$\pm$0.005 & 5.588$\pm$0.003 & 5.367$\pm$0.004\\
        2917.83 & 8.238$\pm$0.002 & 6.039$\pm$0.002 & 8.429$\pm$0.005 & 8.031$\pm$0.006 & 6.161$\pm$0.003 & 5.903$\pm$0.004\\
        2961.81 & 8.913$\pm$0.003 & 6.632$\pm$0.002 & 9.060$\pm$0.005 & 8.740$\pm$0.006 & 6.730$\pm$0.004 & 6.506$\pm$0.004\\
        3009.36 & 9.324$\pm$0.002 & 7.059$\pm$0.002 & 9.479$\pm$0.005 & 9.137$\pm$0.006 & 7.183$\pm$0.004 & 6.911$\pm$0.004\\
        3058.12 & 10.167$\pm$0.003 & 7.703$\pm$0.002 & 10.243$\pm$0.005 & 9.967$\pm$0.006 & 7.765$\pm$0.004 & 7.558$\pm$0.005\\
        3105.20 & 10.858$\pm$0.003 & 8.138$\pm$0.002 & 10.946$\pm$0.006 & 10.576$\pm$0.007 & 8.209$\pm$0.004 & 7.939$\pm$0.005\\
        3151.63 & 11.412$\pm$0.003 & 8.669$\pm$0.002 & 11.565$\pm$0.006 & 11.083$\pm$0.008 & 8.760$\pm$0.004 & 8.458$\pm$0.006\\
        3196.12 & 11.788$\pm$0.003 & 9.157$\pm$0.002 & 11.823$\pm$0.006 & 11.552$\pm$0.008 & 9.176$\pm$0.005 & 9.004$\pm$0.006\\
        3239.26 & 12.651$\pm$0.003 & 9.548$\pm$0.002 & 12.742$\pm$0.006 & 12.366$\pm$0.008 & 9.593$\pm$0.005 & 9.372$\pm$0.006\\
        3281.75 & 12.685$\pm$0.004 & 9.730$\pm$0.003 & 12.720$\pm$0.007 & 12.408$\pm$0.009 & 9.742$\pm$0.005 & 9.558$\pm$0.007\\
        3326.58 & 12.837$\pm$0.003 & 10.046$\pm$0.003 & 12.854$\pm$0.006 & 12.632$\pm$0.008 & 10.042$\pm$0.005 & 9.898$\pm$0.006\\
        3373.55 & 12.715$\pm$0.003 & 10.075$\pm$0.003 & 12.694$\pm$0.006 & 12.469$\pm$0.008 & 10.067$\pm$0.005 & 9.906$\pm$0.006\\
        3418.65 & 13.117$\pm$0.004 & 10.370$\pm$0.003 & 13.053$\pm$0.007 & 12.808$\pm$0.009 & 10.326$\pm$0.006 & 10.164$\pm$0.007\\
        3462.49 & 13.394$\pm$0.004 & 10.425$\pm$0.003 & 13.483$\pm$0.007 & 13.129$\pm$0.009 & 10.493$\pm$0.005 & 10.252$\pm$0.007\\
        3506.96 & 12.910$\pm$0.004 & 10.452$\pm$0.003 & 12.942$\pm$0.007 & 12.620$\pm$0.009 & 10.487$\pm$0.006 & 10.267$\pm$0.007\\
        3555.87 & 13.142$\pm$0.004 & 10.673$\pm$0.003 & 13.181$\pm$0.007 & 12.783$\pm$0.009 & 10.720$\pm$0.006 & 10.432$\pm$0.007\\
        3607.76 & 13.309$\pm$0.004 & 10.939$\pm$0.003 & 13.540$\pm$0.008 & 13.019$\pm$0.009 & 11.123$\pm$0.006 & 10.737$\pm$0.008\\
        3666.86 & 12.817$\pm$0.004 & 10.664$\pm$0.003 & 13.144$\pm$0.008 & 12.606$\pm$0.010 & 10.920$\pm$0.006 & 10.500$\pm$0.008\\
        3736.66 & 11.349$\pm$0.004 & 9.596$\pm$0.003 & 11.687$\pm$0.008 & 11.181$\pm$0.010 & 9.872$\pm$0.007 & 9.460$\pm$0.008\\
        3830.47 & 8.498$\pm$0.003 & 7.385$\pm$0.003 & 8.674$\pm$0.006 & 8.362$\pm$0.008 & 7.533$\pm$0.006 & 7.270$\pm$0.007\\
        3957.56 & 6.383$\pm$0.003 & 5.785$\pm$0.002 & 6.522$\pm$0.005 & 6.317$\pm$0.006 & 5.906$\pm$0.005 & 5.726$\pm$0.006\\
        4125.98 & 4.589$\pm$0.002 & 4.341$\pm$0.002 & 4.709$\pm$0.004 & 4.511$\pm$0.005 & 4.453$\pm$0.004 & 4.269$\pm$0.005\\
        \B 4282.77 & 2.882$\pm$0.003 & 2.765$\pm$0.003 & 2.945$\pm$0.005 & 2.856$\pm$0.007 & 2.825$\pm$0.005 & 2.739$\pm$0.006
		\\
		\hline
	\end{tabular}
\end{table*}
\end{appendix}

\end{document}